\documentclass[journal]{IEEEtran}
\pdfoutput=1
\usepackage[T1]{fontenc}
\usepackage{graphicx}
\usepackage{tabularx}
\usepackage{amsmath}
\usepackage{amsfonts}
\usepackage{amssymb}
\usepackage{amsthm}
\usepackage{mathbbol}
\usepackage{color, colortbl}
\usepackage{caption}
\usepackage{diagbox}
\usepackage{multirow}
\usepackage[hyphens]{url}
\usepackage[capitalize]{cleveref}
\usepackage[shortlabels]{enumitem}
\usepackage{commands}
\usepackage{algorithm}
\usepackage{algorithmic}
\usepackage{cite}
\usepackage{bm}
 
\theoremstyle{plain}
\newtheorem*{assessment}{Assessment}

\begin{document}

\title{Target Speech Extraction: Independent Vector Extraction Guided by Supervised Speaker Identification}

\author{Jiri~Malek,~\IEEEmembership{Member,~IEEE,}
        Jakub~Jansky, Zbynek~Koldovsky, \IEEEmembership{Senior member,~IEEE}, Tomas~Kounovsky, Jaroslav~Cmejla and~Jindrich Zdansky 
\thanks{This work was supported by the Technology Agency of the Czech
Republic (Project No. TO01000027), by The Czech Science Foundation (Project No. 20-17720S) and by the Student Grant Scheme 2021 Project of the Technical University in Liberec.}
\thanks{All the authors were with the Faculty of Mechatronics, Informatics, and Interdisciplinary Studies, Technical University of Liberec,
Studentsk\'a 2, 461 17, Liberec, Czech Republic. e-mail: (jiri.malek(at)tul.cz).}}

\markboth{IEEE Transactions on Audio Speech and Language processing}%
{IEEE Transactions on Audio Speech and Language processing}

\maketitle

\begin{abstract}

This manuscript proposes a novel robust procedure for the extraction of a speaker of interest (SOI) from a mixture of audio sources. The estimation of the SOI is performed via independent vector extraction (IVE). Since the blind IVE cannot distinguish the target source by itself, it is guided towards the SOI via frame-wise speaker identification based on deep learning. Still, an incorrect speaker can be extracted due to guidance failings, especially when processing challenging data. To identify such cases, we propose a criterion for non-intrusively assessing the estimated speaker. It utilizes the same model as the speaker identification, so no additional training is required. When incorrect extraction is detected, we propose a ``deflation'' step in which the incorrect source is subtracted from the mixture and, subsequently, another attempt to extract the SOI is performed. The process is repeated until successful extraction is achieved. The proposed procedure is experimentally tested on artificial and real-world datasets containing challenging phenomena: source movements, reverberation, transient noise, or microphone failures. The method is compared with state-of-the-art blind algorithms as well as with current fully supervised deep learning-based methods. 

\end{abstract}


\IEEEpeerreviewmaketitle

\section{Introduction}
\label{sec:intro}

\IEEEPARstart{A}{} frequent goal of speech processing is to recover a speaker of interest (SOI) from a mixture of speech sources and environmental noise. This task is often solved using \textit{speech separation} methods; all sources present in the mixture are estimated and subsequently the SOI is identified among them. Separation can be performed either via \textit{data-driven techniques} deriving their models from large sets of training signals~\cite{overview:audiosep:supervised,mc-wsj0-2mix,tasnet,mlss:togami:icassp2021,mlss:boedecker:icassp2021} or via \textit{model-based techniques} utilizing general statistical assumptions about the sources and the mixing~\cite{book:audiobss,iva:def,bss:iva:icassp2021,bss:mnmf:icassp2021,bss:ilrma:def,bss:ilrma:advsig2020,bss:ilrma:icassp2021}.

Both approaches have distinct advantages. The data-driven techniques~\cite{overview:audiosep:supervised} employ principles of \textit{machine/deep learning} to estimate the separating models. If provided with relevant scenario-specific training data, they achieve high separation quality. The machine learning-based approaches primarily perform spectral filtering on single-channel data, which can be supplemented by additional spatial filtering if additional channels are available. In contrast, the model-based approaches~\cite{book:audiobss} employ only general statistical models of the sources/mixing. These approaches do not need any training data and require only minimum information about the target scenario, i.e., they perform the \textit{blind separation}. The blind approaches are, in theory, applicable to a wide range of tasks without any need for adaptation. Arguably, this freedom is achieved at the cost of lower separation accuracy since the employed statistical models only approximate the real conditions. Many blind techniques aim to estimate spatial filters and thus require multi-channel mixtures.

Focusing on blind methods, separation of speech usually proceeds in the time-frequency domain. Independent component analysis (ICA, \cite{book:ica}) separates the sources based on their statistical independence. For a wide-band signal, ICA is separately applied to each frequency bin, which leads to the so called \emph{permutation ambiguity}~\cite{permprob:sawada}. The recovered frequency components have a random order and all components corresponding to the wide-band source need to be identified in order to reconstruct it in the time-domain. To alleviate this drawback, the independent vector analysis (IVA, \cite{iva:def,bss:iva:icassp2021}) has been proposed. It binds together the frequency components corresponding to a single source using higher-order dependencies among them. Non-negative matrix factorization (NMF, \cite{bss:nmf}) attempts to factorize spectrogram of a single-channel mixture as a product of two non-negative components, recurring patterns and their activations. Multi-channel NMF (MNMF, \cite{bss:mnmf:icassp2021}) extends this concept for analysis of multi-channel mixtures. Independent low-rank matrix analysis (ILRMA, \cite{bss:ilrma:def,bss:ilrma:advsig2020,bss:ilrma:icassp2021}) unifies the principles of IVA and NMF. Spectral masking methods~\cite{mandel2010,bss:smplx:taslp2020} use the assumption that only one source is dominant at each time-frequency point.

The full separation attempts to estimate all sources in the mixture, which usually requires the knowledge of the number of sources. This arguably limits the practicality/flexibility of the separation (see the discussion in~\cite{speakerbeam:timedomain,mlse:multistage:icassp2021,mlse:adenet:icassp2021}). To alleviate, the recovery can be focused exclusively on the SOI, which is referred to as \textit{target speech extraction}. The extraction can thus be interpreted as simultaneous identification and estimation of the SOI. Again, this task can be solved using machine learning-based approach~\cite{speakerbeam:timedomain,mlse:multistage:icassp2021,speakerbeam:journal,mlse:spkbeamext1:icassp2021,mlse:voicefilter:is2019,mlse:spex:taslp} or blind source extraction (BSE,~\cite{koldovsky2019TSP, bss:ivewpe:letters2021}).

The independent vector extraction (IVE) is a sub-problem of IVA focusing on the SOI~\cite{koldovsky2019TSP,scheibler2019independent,ive:auxi:araki2020,bss:ivewpe:letters2021}. By definition, IVE methods extract an arbitrary source depending on (often random) initialization. To extract the SOI, information identifying this source is required. Such information can be provided via an initialization focused on the SOI. The utilization of a video stream was proposed to this end in~\cite{prior:init}. Such initialization, however, does not guarantee that the method will remain focused on the SOI during the updates. Alternatively, the extraction can be limited to a direction containing the SOI via the geometric constraint~\cite{prior:geometry:trsp2020,bse:spatconst:icassp2020}, which requires information about SOI location. A practical alternative is to introduce a \textit{pilot signal} that is related to the SOI and directs the convergence towards it. For example, piloting using voice activity detection was proposed for mixtures containing a single speaker in~\cite{pilot:video}. For mixtures of multiple speakers, pilots using supervised speaker identification via embeddings~\cite{xvector:speaker,xvector:diary} have recently proven to be very effective in~\cite{loc:bse:xvec}. The embeddings were used to tackle the problem concerning the ambiguity of the SOI in the deep learning-based extractors as well~\cite{speakerbeam:journal, speakerbeam:timedomain}. 

Speaker embeddings are Deep-Neural-Network-based (DNN) features encoding the characteristics of a speaker. Several variants have recently been introduced, differing mainly in the architecture of the extracting DNN. Embeddings derived from fully-connected feed-forward DNN were proposed in~\cite{emb:dnn}. Approaches utilizing the context of the data via recursive long short-term memory (LSTM) networks were presented in~\cite{emb:lstm}. The recursive modeling allows for more precise classification, however, the training is data-demanding and time-consuming. 

To alleviate those demands, non-recursive architectures capturing the context have been proposed, such as time-delayed neural networks~\cite{tdnn:intro} (TDNN) or feed-forward sequential memory networks~\cite{emb:fsmn} (FSMN). The ``context layers'' within these networks process a set of frames (or feature vectors produced by their previous layer) centered around the current frame. The processing of the context significantly increases the number of learnable parameters. To reduce this number, TDNN sub-samples the set of frames, because the neighboring frames are assumed to be correlated. In contrast, FSMN weights all frames at an input of a layer by a trainable matrix and performs mean time-pooling. Thus FSMN can be seen as a generalization of TDNN in which the importance of frames is learned during training rather than selected during design. 

The traditional model of IVE is time invariant, i.e., it is suitable for separation of immobile sources (\textit{static approach}). To extract moving sources, the time invariant methods are consecutively applied to short intervals of data where the sources are approximately static, and their parameters are recursively updated. The drawback of this \emph{block-wise static approach}~\cite{loc:bse:xvec} lies in difficult tuning of the block length and the recursion weight. Recently, an alternative approach based on the constant separating vector (CSV) model~\cite{loc:auxive} has been proposed. It allows for changes of mixing parameters within the processed interval of data. Compared to the block-wise static approach, the CSV-based method exploits longer intervals of signals. Consequently, this allows us to achieve a higher extraction accuracy. This higher accuracy was proven theoretically in~\cite{csv:cramer:rao} and demonstrated experimentally in~\cite{loc:bse:xrob}.

In this manuscript, we propose an improvement of a guided blind IVE-based method involving the above described advancements introduced in our previous works~\cite{loc:auxive,loc:bse:xvec,loc:bse:xrob}. It consists of a combination of model-based extraction and data-driven frame-wise identification of the SOI. The extraction is based on an IVE algorithm endowed with the CSV mixing model. The IVE algorithm is guided towards the SOI using piloting exploiting speaker embeddings computed via an FSMN model. This combination simplifies/reduces the amount of training compared to fully data-driven techniques. The proposed method needs only to learn how to identify the SOI while its extraction is free of training, which makes it applicable to a wide variety of realistic mixtures. Thanks to this decoupling of the identification and the extraction, the identification can be trained generally, independent of a specific mixing scenario.

The contribution of this manuscript to the basic concept is threefold. 1) An improvement for piloting is introduced by incorporating a non-intrusive criterion for the assessment of the extraction performance. The assessment allows for the detection of the cases in which an incorrect source is being extracted. 2) These incorrect cases are treated using a deflation approach: the unwanted source is subtracted from the mixture, and the extraction is attempted again. This cycle continues until the SOI is extracted. The pilot signal and the criterion share the same pretrained FSMN model, i.e., no additional training is required. 3) Compared to our previous works, we perform a more detailed experimental analysis of the properties and limitations of piloting. The proposed extractor is verified using two widely analyzed datasets (CHiME-4~\cite{chime:web}, the spatialized version of wsj0-2mix~\cite{mc-wsj0-2mix}) and an ad-hoc dataset featuring source movements. The benefits of the deflation step are demonstrated, and the results are compared to the state-of-the-art deep-learning-based and blind methods.

This article is organized as follows. The blind extraction algorithm is described in Section~\ref{sec:bse}. Section~\ref{sec:pilot} provides the principles of piloting. The non-intrusive criterion for assessment of extraction quality is proposed in Section~\ref{sec:asses}. The deflation is presented in Section~\ref{sec:defl}. The proposed method is experimentally evaluated in Section~\ref{sec:exp}, while Section~\ref{sec:concl} concludes the manuscript.

\section{Algorithm description}
\label{sec:desc}

\subsection{Problem definition}
\label{sec:problem_description}

A time varying mixture of $d$ original signals observed by $d$ microphones can, in the short-time frequency domain, be approximated by the mixing model
\begin{equation}
\x = \Al \y,
\label{eq:model}
\end{equation}
where $k=1,\dots,K$ is the frequency and $\ell=1,\dots,L$ is the frame index. $\x \in \mathbb{C}^d$ denotes a vector of the mixed signals recorded on $d$ microphones, $\y \in \mathbb{C}^d$ is a vector whose $i$th component corresponds to the $i$th original signal and $\Al \in \mathbb{C}^{d \times d}$ is the mixing matrix. For practical reasons, it is often assumed that the mixing is approximately static over a small number of subsequent frames. Let this interval be referred to as \textit{block} in this manuscript. The mixture is thus divided into $t = 1,\dots,T$ equally long blocks of length $L_T$ frames with a block-constant mixing matrix $\A$; the index of the $\ell$th frame within the $t$th block is denoted by $\lt$. The mixing model for this \emph{block-wise static} approach \cite{loc:bse:xvec} is given by
\begin{equation}
\xt = \A \yt \qquad \text{for } \ell = 1, \dots,L_T;\ t = 1,\dots,T.
\label{eq:model:bw}
\end{equation}
Note that this model becomes fully static when $T = 1$ or maximally time-varying if $T = L$.

In IVA, a complete de-mixing matrix $\W \in \mathbb{C}^{d \times d}$ is sought such that it fulfills $\W\xt = \W\A \yt = \yth \approx \yt$, i.e., it recovers all the sources present in the mixture. In contrast, IVE seeks only one row of $\W$, denoted by $\wt$, such that it specifically extracts the SOI. Without any loss on generality, let the SOI be the first signal in $\yt$ and $\A$ be partitioned as $ \A = 
\begin{bmatrix}
\at & {\bf Q}_{t}^k
\end{bmatrix}$. Then, the equation \eqref{eq:model:bw} can be expressed in the form
\begin{equation}
\xt = [\at \quad {\bf Q}_{t}^k]\begin{bmatrix} \st \\
\bzltk \\
\end{bmatrix},  
\end{equation}
where $\st$ represents the SOI and $\bzltk$ are the other $d-1$ signals in the mixture. Subsequently, $\W$ can be partitioned as $[\wt \quad ({\bf B}_{t}^k)^H]^H$, where, ${\bf B}_{t}^k$ is called a {\it blocking matrix} and $\cdot^H$ denotes conjugate transpose.

\subsection{Blind extraction: CSV-AuxIVE algorithm}
\label{sec:bse}

The extraction part of the proposed procedure is a blind algorithm from~\cite{loc:auxive}. Here, we overview its most important ideas and provide some intuition on how the final update rules were obtained. The CSV model is based on the assumption that the separating vector $\wt$ is constant within all $T$ blocks ($\wt=\w, t=1\dots T$). This means that the separating vector obeys $(\w)^H\xt= \ut \approx \st$ for each block $t$, where $\ut$ is the SOI estimate. The mixing vector $\at$ and the blocking matrix ${\bf B}_{t}^k$ are still assumed to vary with respect to $t$.

The estimation of $\w$ stems from the following \textit{log-likelihood function}. Let $\bslt = [\slt^1 \dots \slt^K]$ be a vector of all frequency components corresponding to the SOI. The elements of $\bslt$ are assumed to be dependent; they thus need to be modeled by a joint pdf $p_s(\bslt)$. The background signals $\bzlt^1 \dots \bzlt^K$ are Gaussian and their frequency components are assumed to be uncorrelated. Consequently, their higher order dependencies are zero and they can be modeled as independent; let their density be denoted $p_{{\bf z}}(\bzltk)$. The log-likelihood function is then
\begin{multline}\label{eq:logogive_w}
    \mathcal{L}(\{\w \}_{k \leq K},\{\at \}_{k \leq K}|\{\xt\}_{k \leq K}) =\log p_s(\{\ut\}_{k \leq K})  \\ +\sum_{k=1}^{K} \log p_{{\bf z}}(\zth) +\log |\det \W|^2,
\end{multline}
where $\zth$ is the estimate of the background signals. The notation $\{\cdot\}_{k \leq K}$ describes a variable with all values of index $k$, e.g., $\{\w \}_{k \leq K} = \mathbf{w}^1,\dots,\mathbf{w}^K$. 

Subsequently, \textit{a contrast function} is formulated using the assumption that all samples are independently distributed and the log-likelihood function~\eqref{eq:logogive_w} thus can be averaged over all blocks and samples. Optimization of the contrast function is performed using the \textit{auxiliary function optimization} technique~\cite{ono2011stable}. The main idea is to replace the nonlinear contrast function with an auxiliary function, which is easier to optimize and retains the same optimal solution. Then the new auxiliary function is alternately optimized in the original and the auxiliary variables. Moreover, since the true model of the $p_s(\bslt)$ is unknown, a surrogate density function suitable for speech signals is chosen in the form $f(x) \propto \exp\{-\|x\|\}$. 

The \textit{update rules} for finding the optimum point of the auxiliary contrast function are obtained in the form: 
\begin{align}
    \rt &=\sqrt{\sum_{k=1}^K |(\w)^H\xt|^2} \qquad \text{for all } \lt \label{eq:blockauxiverule1}, \\
    \V &= \hat{\rm E}_t\left[\varphi(\rt)\xt(\xt)^H\right] \label{eq:blockauxiverule2},\\
    \wCxkt &= \hat{\rm E}_t\left[\xt(\xt)^H\right],\label{eq:blockauxiverule3}\\
\at & =\frac{\wCxkt\w}{(\w)^H\wCxkt\w} \label{eq:blockauxiverule4},\\
\hat{\sigma}_{k,t} &= \sqrt{(\w)^H\wCxkt\w}, \label{eq:blockauxiverule5}\\
\w &\leftarrow  \left(\sum_{t=1}^{T}\frac{\V}{(\sigh)^2}\right)^{-1} \sum_{t=1}^{T}\frac{(\w)^H \V\w}{(\sigh)^2}\at   \label{eq:blockauxiverule6},
\end{align}
where $\rt$, $\V$ are the auxiliary variables, $\varphi(\rt) = \rt^{-1}$ is a nonlinearity suitable for super-Gaussian signals such as speech, $\Ct$ is the sample-based covariance matrix of the mixture on the $t$th block and $\Eth$ denotes the sample-based expectation over the frames in block $t$. Equation~\eqref{eq:blockauxiverule4} is the \textit{orthogonal constraint} (OGC) ensuring mutual orthogonality of subspaces generated by the SOI and the other signals and $\sigh$ is the sample-based variance of the SOI. A normalization $\w \leftarrow \w/\sqrt{\sum_{t=1}^{T}(\w)^H \V\w}$ is performed after each iteration (i.e., sequence of update rules \eqref{eq:blockauxiverule1}--\eqref{eq:blockauxiverule6}) to enable stable convergence.

When $T = 1$,  CSV-AuxIVE corresponds to the auxiliary function-based IVE for  static sources from~\cite{scheibler2019independent}, which is denoted as FS-IVE in the experiments within Section~\ref{sec:exp}. Successive application of the static IVE to blocks $t=1\dots T$ gives the block-wise static IVE approach from \cite{loc:bse:xrob} (BS-IVE) that allows for dynamic mixing. To cope with the lack of data in short blocks, BS-IVE performs the following two steps. First, the extraction on the block $t$ is initialized by the de-mixing vectors achieved on the block $t-1$, and, second, the statistics required in the update rules \eqref{eq:blockauxiverule1}--\eqref{eq:blockauxiverule6} are computed in a recursive manner.

\subsection{Extraction guided towards the SOI: piloting using the supervised speaker identification}

CSV-AuxIVE extracts an arbitrary source from the mixture if no prior information concerning the SOI is available. This section discusses how the supervised speaker identification via embeddings is used to focus the extraction on the SOI. First, our implementation of the FSMN network for computation of the conventional \textit{sentence-wise embeddings} is described. Subsequently, a general concept of a pilot signal is introduced. The pilot signal is statistically dependent on the SOI. It is submitted to the CSV-AuxIVE with the mixture and forces the blind algorithm to converge towards the SOI. Finally, modifications to the FSMN network are proposed, which allow computation of \textit{frame-wise embeddings} and the design of a practically usable pilot.

\subsubsection{Network producing the embeddings, X-vectors}
\label{sec:tdnn}

Our implementation of the embedding network stems from the FSMN\footnote{The utilized network architecture does not differ from our previous works in~\cite{loc:bse:xvec,loc:bse:xrob}. There we described the embedding network as TDNN with modifications. A more detailed research of literature revealed that it is more accurate to label the network as FSMN.} architecture~\cite{emb:fsmn} and is summarized in Table~\ref{tab:xvec:net}. Its input consists of a single-channel audio signal sampled at $16$~kHz. The input features are $40$ filter bank coefficients computed from frames of a length of $400$ and a frame-shift of $200$ samples. Subsequently, six Context layers are present, i.e., context of frames is weighted by a trainable matrix; mean time-pooling is performed; and a linear transformation is applied. The output of each layer is weighted by the exponential linear unit (ELU). The Pooling layer computes variances of frames. Its context length is $L_c=101$ during training. Overall, the size of the model is $1.8$ million parameters. The network is trained to classify $N$ speakers via minimization of the cross-entropy loss function.

After training, the two latest classification layers are removed and the embeddings are extracted from the Pooling layer. This is done to allow for classification of the speakers absent in the training set. In the test phase, an embedding of an unknown speaker is compared to the set of embeddings (called \textit{enrollment}) corresponding to the potential speakers. This comparison is performed by Probabilistic Linear Discriminant Analysis (PLDA,~\cite{plda}). PLDA is a machine learning approach that tests a hypothesis that a an enrollment vector $\enr$ and test vector $\rgen$  corresponds to a single speaker. The statistical distributions necessary for this testing are derived from a training dataset of precomputed embeddings. PLDA returns a score $M(\enr,\rgen)$, which is high if the hypothesis is correct. 

The training data for the FSMN and PLDA originate from the development part of the Voxceleb1 database~\cite{voxceleb} and the training part of the LibriSpeech corpus~\cite{librispeech}. The recording of Voxceleb1 ($149$k utterances, about $340$ hours) contain real-world reverberation and noise. Librispeech (part train-360-clean, $104$k utterances, $360$ hours) is free of distortions. It was subjected to augmentations discussed below, in order to train X-vectors robust with respect to environmental distortions. The environmental noise was taken from the simulated part of the CHiME-4 training dataset~\cite{chime:web} and the development dataset available in Task 1 of the DCASE2018 challenge~\cite{dcase:web}.

The augmented X-vectors were trained on one unchanged instance of Voxceleb1/Librispeech and three augmented instances of the Librispeech dataset, where the following augmentations were applied:
\begin{enumerate}
\item Reverberation: The utterances were convolved with artificial room impulse responses (RIRs) generated by~\cite{rir:gen:habets}. The artificial RIRs originated from a shoe-box room of size $8 \times 7 \times 3$~m; four different reverberation times $T_{60}$, ranging from $175 - 650$~ms, were considered. The source-microphone distance was $1-2$~m. 
\item Noise: The environmental noise was summed with the original Librispeech utterances at signal-to-noise-ratio (SNR) equal to $10$~dB.
\item Reverberation+noise: The noise was added to the reverberated Librispeech dataset with SNR$=10$~dB. 
\end{enumerate}
The PLDA was trained using the three augmented variants of the Librispeech dataset.

In this manuscript, we denote the extracted embeddings as X-vectors. In a narrow sense, this term is reserved for features estimated by the TDNN~\cite{xvector:speaker}. However, since both topologies are closely related, we believe such naming can be used without ambiguity.

\begin{table}
\centering 
\caption{\label{tab:xvec:net} Description of the FSMN producing the X-vectors. The input sizes for the context layers are stated after the mean pooling operation.}
\begin{tabular}{|c|c|c|c|}
 \hline
 \textbf{Layer} & \textbf{Layer} & \textbf{Total} & \textbf{Input} \\
  & \textbf{context} & \textbf{context} & x \textbf{output} \\ 
 \hline\hline
Context 1  & $\ell\pm80$ & $161$ & $40 \times 1024$ \\ \hline
Context 2  & $\ell\pm4$ & $169$ & $1024 \times 768$ \\ \hline
Context 3  & $\ell\pm4$ & $177$ & $768 \times 512$ \\ \hline
Context 4  & $\ell\pm4$ & $185$ & $512 \times 384$ \\ \hline
Context 5  & $\ell\pm4$ & $193$ & $384 \times 256$ \\ \hline
Context 6  & $\ell\pm4$ & $201$ & $256 \times 128$ \\ \hline
Fully-conn. 1 & $\ell$ & $201$ & $128 \times 128$ \\ \hline
Pooling & $\ell\pm \frac{\Lc-1}{2}$ & $201+\Lc$ & $(\Lc\cdot128)\times 128$ \\ \hline
Fully-conn. 2  & $\ell$ & $201+\Lc$ & $128\times 128$ \\ \hline
Softmax  & $-$ & $201+\Lc$ & $128\times N$ \\ \hline
\end{tabular}
\end{table}

\subsubsection{The concept of piloting}
\label{sec:pilot}

The pilot signal represents an information identifying the SOI for the CSV-AuxIVE. It forces the blind algorithm to converge towards the SOI. The pilot signal is introduced through modification of the update step in~\eqref{eq:blockauxiverule1}. This equation corresponds to a factor that binds together all frequency components belonging to a single source. Without this factor, the independence of the outputs would be achieved in each frequency bin $k$ separately, and the reconstruction of the wide-band SOI would suffer the permutation problem described in the Introduction. Modification of the equation~\eqref{eq:blockauxiverule1} into the form
\begin{equation}\label{eq:auxivepiloted}
\rt=\sqrt{\sum\nolimits_{k=1}^K |(\w)^H \xt|^2 + \plt}.
\end{equation}
adds the dependency of all the frequency components on the pilot signal $\Pilot$ and consequently also the SOI. The pilot $\Pilot$ is independent of the mixing model parameters and thus does not change the remaining update rules of the CSV-AuxIVE.

The signal $\Pilot$ needs to be designed as statistically dependent on the SOI. The term under the square root of~\eqref{eq:blockauxiverule1} describes the total energy of the extracted components. Thus the frame-wise energy of the SOI appears to be a suitable choice. Since the actual energy is unknown and difficult to estimate, a reasonable approximation is given by the frames of the mixture, where the energy of the unwanted sources is low (the energy of the SOI is \textit{dominant}). We propose to compute the pilot signal $\Pilot$ for the $\ell$th frame (note that $\Pilot$ is independent of CSV blocks) as 
\begin{equation}\label{eq:pilot}
    \pl =\begin{cases}
    \sum_{k=1}^K|\xp|^2 & \text{the SOI is dominant},\\
    0& \text{otherwise,}\\
    \end{cases}
\end{equation}
where $\xp$ is the mixture on the first microphone. Specific pilot signals (and their respective ways how to determine the dominance of the SOI) are introduced in Section~\ref{sec:xvec:pilot}. 

\subsubsection{Frame-wise speaker identification for piloting}
\label{sec:xvec:spkid}

The utilization of X-vectors and PLDA for piloting differs from the conventional speaker identification in several aspects.
\begin{enumerate}[a)]
\item \label{itm:soiid:frmwise} Conventionally, speaker identification operates on long intervals/sentence-wise. However, the pilot signal in~\eqref{eq:pilot} requires frame-wise information about the dominance of the SOI, i.e., \textit{a frame-wise sequence of X-vectors}. Each X-vector then describes the identity of a speaker in a short interval centered around the current frame. To obtain such a sequence, the input context of the FSMN is gradually shifted by a single frame. For each shift, an X-vector is computed based on pooling with a shortened context (e.g., $L_c=11$).
\item \label{itm:soiid:crstlk} The identification is performed in the presence of cross-talk. However, only the identity of \textit{the dominant speaker} is required (due to the definition of the pilot in~\eqref{eq:pilot}).
\item \label{itm:soiid:soi} Only the \textit{identity of the SOI is of interest}; it must not be confused with any interfering speaker. The substitution within the set of interferers is irrelevant because the pilot in~\eqref{eq:pilot} is set to zero when any unwanted source is assumed dominant.
\item \label{itm:soiid:enrnum} The set of the potential speakers (enrollment set) is significantly smaller. Conventionally, hundreds of speakers need to be distinguished. For the purposes of piloting, the enrollment set contains X-vector for each speaker, which can be active in the processed dataset (e.g., $18$~vectors for the wsj0-2mix dataset~\cite{mc-wsj0-2mix}). 
\end{enumerate}
The aspects \ref{itm:soiid:frmwise} and \ref{itm:soiid:crstlk} complicate the identification task, whereas the aspects~\ref{itm:soiid:soi} and~\ref{itm:soiid:enrnum} simplify it. 

To perform the frame-wise identification, PLDA scores $M_\ell(\enr,\rXl)$ are computed. Here, $\enr$ is the enrollment X-vector corresponding to one of the potential speakers and $\rXl$ is the X-vector computed using the context around the $\ell$th frame within the first channel of the mixture. The speaker with the highest $M_\ell(\enr,\rXl)$ is the most distinctive from the perspective of the X-vectors and is also \textit{assumed to be dominant} in the mixture. Validity of this assumption is experimentally verified in Section~\ref{sec:case:xacc}. 

\subsubsection{Specific pilot variants}
\label{sec:xvec:pilot}

Two variants of a pilot signal are considered in this manuscript. The properties and limitations of the proposed pilots are demonstrated experimentally in Section~\ref{sec:exp:pilot}.

The realizable \textit{X-vector-based pilot} $\PilotX$ is computed according to~\eqref{eq:pilot}, where the SOI is considered dominant in the $\ell$th frame if 
\begin{align}
    \label{eq:dom:xvec}
    M_\ell(\enrS,\rXl) &> \text{max} \{M_\ell({\enrZj},\rXl), j=1\dots J\} \ \ \text {and} \nonumber \\
    M_\ell(\enrS,\rXl) &> \pldaMin(\enrS),
\end{align}
where $\enrS$ denotes the X-vector corresponding to the SOI and $\pldaMin(\enrS)$ is the lowest PLDA score, where the SOI is still considered active. The variable ${\enrZj}$ denotes the X-vector corresponding to the $j$th potential interfering speaker from the enrollment set containing the SOI and $J$ other speakers. To compute the $\PilotX$, the following two pieces of information are \textit{needed}: the identity of SOI (we need to know which source we want to extract) and the enrollment set containing the X-vector for each speaker present in the processed dataset. On the other hand, the number of sources in the mixture or the identities of the active interferers are \textit{not required}.

An \textit{oracle pilot} $\PilotO$ is used to analyze the possibilities of the piloting proposed via~\eqref{eq:pilot}. The dominance of the SOI is always determined correctly using true unobservable energies of the sources. Due to the use of unavailable information, it cannot be used in practice. $\PilotO$ is computed using~\eqref{eq:pilot}, where SOI is considered dominant within the $\ell$th frame if 
\begin{equation}\label{eq:dominant_orac}
    \sum_{k=1}^K|\slk|^2 > \thro\sum_{k=1}^K||\zl||^2,
\end{equation}
where $\thro$ is a free parameter reflecting the desired level of dominance.

\subsection{Non-intrusive assessment of extraction quality}
\label{sec:asses}

This Section proposes a non-intrusive criterion to assess whether the extraction of the SOI was successful. This criterion is based on the same X-vectors and PLDA as the piloting, i.e., no additional training is required. 

The assessment represents the entire signal through a single PLDA score $M(\enrS,\hrSf)$, where $\hrSf$ is the X-vector independent of $\ell$ computed from an estimate of the SOI (FSMN pooling context is set $L_c=L$). As in the conventional speaker identification, this score can be seen as a measure of similarity between the X-vector computed from the enrollment utterance of the SOI and an unknown test X-vector.
The two following observations concerning the score hold. 1) When the SOI is truly active in the test utterance, its PLDA score is higher than the non-active speakers' scores. 2) Interferences decrease the similarity/score compared to values observed on undistorted test signals. Based on these observations, the \textit{extraction assessment} is proposed:
\begin{assessment}[of extraction quality]
    \label{prop:quality}
    Having X-vectors for two signals containing the same component corresponding to the SOI\footnote{For example, the original mixture and the extracted signal.} denoted by $\hrSf,\hrSs$; if $M(\enrS,\hrSf) > M(\enrS,\hrSs)$ then $\hrSf$ corresponds to a superior estimate of this SOI component in the sense of speech quality. 
\end{assessment}
The extraction assessment is experimentally validated in Section~\ref{sec:exp:asses}. The Section shows a strong linear dependence between increments in criteria measuring quality of speech and the PLDA score. The extraction assessment is used in the deflation process as a \textit{decision mechanism}. It determines whether the extracted source is an estimate of the SOI or of an unwanted source (and the deflation should be applied).

\subsection{Re-estimation of the SOI on extraction failure: deflation}
\label{sec:defl}

\begin{algorithm}[t]
		\caption{\label{alg:defl}Deflation mechanism for CSV-AuxIVE using the extraction assessment. Variables $\hrS^i$ and $\rX^i$ denote X-vectors corresponding to the SOI estimate and the first channel of the mixture after $i$ deflation steps.}
		\begin{algorithmic}
			\REQUIRE {Multi-channel mixture ${\bf x}^k_\ell$, X-vector FSMN, enrollment set including the SOI, PLDA model}
		    \FOR{$i \leftarrow 0, i < I$}
                \STATE{Extract $\shkit$ from $\xkit$ using piloted CSV-AuxIVE}
                \IF{$M(\enrS,\hrS^i)$ > $M(\enrS,\rX^i)$}
                    \RETURN{$\shkit$} \COMMENT{Extracted source is the SOI estimate} 
                \ELSE
                    \STATE{$\xkiot \leftarrow$ Subtract $\shkit$ from $\xkit$ using~\eqref{eq:defl:subtr}}
                    \IF{$M(\enrS,\rX^i)$ > $M(\enrS,\rX^{i+1})$}
                        \RETURN{$\xkit$} \COMMENT{Reduced mixture is not closer to the SOI, end the deflation}
                    \ELSE
                        \STATE{} \COMMENT{Continue the deflation}
                    \ENDIF
                \ENDIF
    	    \ENDFOR
    	    \RETURN{$\xkiot$} \COMMENT{Maximum number of steps reached}
		\end{algorithmic}
\end{algorithm}

The deflation provides a mechanism to extract the SOI from mixtures in which the desired source is difficult to identify via pilot alone. This may happen, e.g., when the SOI is the weaker source and only a small number of frames with dominant SOI exist to form an efficient pilot. 

The deflation is summarized in Algorithm~\ref{alg:defl} and proceeds as follows. The first signal is extracted using the piloted CSV-AuxIVE. The extraction assessment is used to determine whether this signal represents a better estimate of the SOI than the original mixture. If so, the first signal is returned and the extraction ends. Otherwise, the first signal is subtracted from the mixture (on each CSV block) using least square projection. Using the assessment, the reduced mixture is compared to the original one. If the original mixture is chosen, the extraction ends (the deflation did not bring the mixture closer to the SOI). If the reduced mixture is selected, the piloted CSV-AuxIVE is applied to it and the second signal is extracted. This process is repeated until an estimate of the SOI is found or until a predefined number $I$ of deflation steps has been performed. It is reasonable to select $I$ close to the assumed number of speakers active in the mixture. Owing to the utilization of the pilot signal, the CSV-AuxIVE is forced to converge towards speech signals. Thus, the active speakers are the first extracted sources in most cases.

Let $\xkit \in \mathbb{C}^{d-i}$ and $\wki \in \mathbb{C}^{d-i}$ denote the input mixture and the separating vector after $i$ deflation steps, respectively. The reduced mixture $\xkiot \in \mathbb{C}^{d-i-1}$ is obtained by the least square subtraction of the extracted signal $\shkit=(\wki)^H\xkit$ from $\xkit$. Let $\akit$ be the mixing vector after $i$ deflation steps computed on the $t$th block via~\eqref{eq:blockauxiverule4}. Due to the orthogonality of $\wki$ and $\akit$, the subtraction is achieved through
\begin{equation}
    \label{eq:defl:subtr}
    \xkiot = \bDki(\xkit -\akit(\wki)^H\xkit),
\end{equation}
where $\bDki$ is a $(d-i-1)\times(d-i)$ full row-rank matrix, reducing the dimension of $\xkiot$ by one compared to $\xkit$. This reduction needs to be applied to avoid rank deficiency of the ``deflated'' mixture. Matrix $\bDki$ can be found via principal component analysis~\cite{pca} or can simply omit one element of $\xkit$.

\section{Experiments}
\label{sec:exp}

The following experiments pursue three goals. 1) The possibilities and limitations of piloting are investigated as a motivation for the proposed deflation. 2) The functionality of the proposed extraction assessment is analyzed. 3) The benefits of the deflation are demonstrated and the performance of the proposed extractor is compared to results published in the literature. 

\subsection{Datasets}

The experiments are performed on the following three datasets, which contain various detrimental phenomena such as source movements, high reverberation and noise activity, transients or low energy of the SOI.

\subsubsection{Dynamic dataset}
\label{sec:data:case}

The first dataset is an ad-hoc simulated one containing noisy recordings of two simultaneously active moving speakers (SOI and an interfering source (IS)). The sources are located in a room of dimensions $6 \times 6 \times 3$ m; reverberation times $T_{60} \in \{100,300,600\}$~ms are considered. A linear array of five omni-directional microphones with spacing of $8$~cm is placed close to the center of the room and rotated counter-clockwise by $45^{\circ}$. Both sources move on a half-circle around the array, the radius is $1.5$~m for the SOI and $2$~m for the IS. SOI performs a large angular movement in the left-hand half-plane and IS a small one in the right-hand half-plane. A static directional noise source is located perpendicular to the microphone array axis to the right. The situation is depicted in Fig.~\ref{fig:base:movs}.

The speech (sampled at $16$~kHz) originates from the test/development sets of CHiME-4; four potential speakers (F01, F06, M04, M05) are considered.  The cafeteria sounds used for a directional noise originates from the QUT corpus~\cite{data:qut:noise}. Different utterances are concatenated to form $5$ unique test signals of length $25$~s for each speaker. The movements of SOI and positions of the static sources are simulated using the RIR generator \cite{rir:gen:habets}.  One instance of the experiment (for one $T_{60}$ value) thus consists of $300$~mixtures ($6$ speaker combinations $\times$ $2$ speaker roles $\times$ $25$ utterance combinations). The sources are mixed at an input signal-to-interference-ratio of $0$~dB (SIR, ratio of energy of SOI and IS) and an input signal-to-noise-ratio of $10$~dB (SNR, ratio of all speech to noise energy).

\subsubsection{CHiME-4 dataset}
\label{sec:data:chime4}

CHiME-4 dataset~\cite{chime:web} contains six-channel real-world and simulated recordings of a single speaker active in a highly noisy environment. The dataset does not contain cross-talk; however, the real-world part contains a lot of microphone failures and transient noises. These non-speech signals are occasionally extracted instead of SOI. 

\subsubsection{Multi-channel Wall Street Journal - 2mix dataset}
\label{sec:data:mcwsj2}

The multi-channel version of the Wall Street Journal - 2mix dataset (MC-WSJ0-2mix, \cite{mc-wsj0-2mix}) is currently often used to compare speaker separation and extraction algorithms. The MC-WSJ0-2mix dataset contains $3,000$ simulated mixtures recorded in a reverberant environment using a microphone array containing eight microphones. Each mixture contains two active speakers, i.e., there is $6,000$ extraction experiments in total. The sources are mixed with SIR between $\langle-5,+5\rangle$~dB. Some of the recordings are very short; their durations range from $1.6$~s to $13.9$~s. The recordings are highly reverberant ($T_{60} \in \langle200,600\rangle$~ms), and captured in rooms with variable dimensions. The geometry of the microphone array is varying, as well as the source-microphone distance, which is $1.3$~m with $0.4$~m standard deviation. The dataset does not contain environmental noise or source movements. The $8$~kHz variant of the mixtures is used\footnote{We interpolate the mixtures to $16$~kHz in order to be able to process it via the FSMN network. We found that this approach gives comparable results to retraining the network on training datasets down-sampled to $8$~kHz.}.

\subsection{Evaluation measures and common settings}

\begin{table*}[t]
\centering 
\setlength{\tabcolsep}{2.2pt}
\caption{Dynamic dataset: the extraction performance of the CSV-AuxIVE (CSV), the static (FS-IVE) and the block-wise static (BS-IVE) IVE techniques.}
\label{tab:exp:base}
\begin{tabular}{|ccc|ccc|ccc|ccc|ccc|}\hline

\multicolumn{15}{|c|}{\textbf{Unprocessed mixture}} \\ \hline
\multicolumn{3}{|c|}{\multirow{3}{*}{}} & \multicolumn{3}{c|}{\bf input PESQ [-]}&\multicolumn{3}{c|}{\bf input SDR [dB]}&\multicolumn{3}{c|}{\bf input SIR [dB]}&\multicolumn{3}{c|}{}\\ \cline{4-12}
 &  &  &100ms&300ms&600ms &100ms&300ms&600ms &100ms&300ms&600ms & & &\\ \cline{4-12} 
  &  &  & 1.15 & 1.20 & 1.27 & 1.10 & 1.12 & 1.14 & 1.10 & 1.12 & 1.14 & & & \\ \hline \hline 

\multicolumn{15}{|c|}{\textbf{Processed using suitable block length}} \\ \hline
\multirow{2}{*}{\bf Method} & \multirow{2}{*}{\bf $L_T$} & \multirow{2}{*}{\bf Pilot}&\multicolumn{3}{c|}{\bf iPESQ [-]}&\multicolumn{3}{c|}{\bf iSDR [dB]}&\multicolumn{3}{c|}{\bf iSIR [dB]}&\multicolumn{3}{c|}{\bf Attenuation}\\ \cline{4-15} 
 &  &  &100ms&300ms&600ms &100ms&300ms&600ms &100ms&300ms&600ms &100ms&300ms&600ms\\ \hline 
$\text{FS-IVE}$ & $2000$ &  \multicolumn{1}{c|}{-} & 0.32 & 0.05 & -0.01 & 6.84 & -0.95 & -3.73 & 15.33 & 6.97 & 4.04 & 0.42 & 0.17 & 0.10\\ $\text{FS-IVE}$ & $2000$ &  $\PilotO$ & 0.61 & 0.22 & 0.11 & 8.05 & 2.74 & -0.17 & 19.72 & 13.31 & 9.72 & 0.32 & 0.19 & 0.12\\
$\text{FS-IVE}$ & $2000$ &  $\PilotX$ & 0.51 & 0.14 & 0.04 & 7.29 & 1.31 & -2.37 & 18.46 & 11.16 & 6.61 & 0.33 & 0.18 & 0.10\\ \hline

$\text{Proposed CSV}$ & $200$ &  \multicolumn{1}{c|}{-} & 0.49 & 0.11 & 0.02 & 7.89 & 0.21 & -2.88 & 17.03 & 7.60 & 4.60 & 0.36 & 0.15 & 0.10\\ 
$\text{Proposed CSV}$ & $200$ &  $\PilotO$ & 0.87 & 0.27 & 0.13 & 11.52 & 3.58 & 0.16 & 22.01 & 13.29 & 9.64 & 0.25 & 0.13 & 0.10\\ 
$\text{Proposed CSV}$ & $200$ &  $\PilotX$ & 0.76 & 0.20 & 0.06 & 10.24 & 2.23 & -1.80 & 20.46 & 11.38 & 6.85 & 0.28 & 0.14 & 0.10\\\hline 
$\text{BS-IVE}$ & $200$ & \multicolumn{1}{c|}{-} & 0.09 & 0.00 & -0.07 & 2.99 & -1.07 & -3.39 & 12.50 & 6.53 & 4.17 & 0.32 & 0.18 & 0.14\\ 
$\text{BS-IVE}$ & $200$ &  $\PilotO$ & 0.39 & 0.16 & 0.05 & 8.11 & 3.54 & 0.47 & 19.34 & 13.54 & 9.92 & 0.26 & 0.17 & 0.14\\ 
$\text{BS-IVE}$ & $200$ &  $\PilotX$ & 0.22 & 0.04 & -0.05 & 5.33 & 0.61 & -2.49 & 15.69 & 9.25 & 5.69 & 0.28 & 0.17 & 0.13\\\hline\hline 
\multicolumn{15}{|c|}{\textbf{Processed using excessively long/short blocks}} \\ \hline 
$\text{Proposed CSV}$ & $800$ &  $\PilotX$ & 0.52 & 0.15 & 0.03 & 7.05 & 1.11 & -2.53 & 18.11 & 10.68 & 6.17 & 0.32 & 0.17 & 0.10\\ 
$\text{BS-IVE}$ & $800$ &  $\PilotX$ & 0.42 & 0.11 & 0.00 & 6.67 & 1.24 & -2.22 & 18.11 & 10.35 & 5.91 & 0.31 & 0.19 & 0.13\\
$\text{Proposed CSV}$ & $50$ &  $\PilotX$ & 0.60 & 0.14 & 0.02 & 10.47 & 1.80 & -1.83 & 17.63 & 8.47 & 4.45 & 0.16 & 0.11 & 0.10\\
$\text{BS-IVE}$ & $50$ &  $\PilotX$ & 0.04 & -0.03 & -0.10 & 2.77 & -0.47 & -3.01 & 11.83 & 7.63 & 4.85 & 0.23 & 0.15 & 0.12\\\hline 

\end{tabular}
\end{table*}

The extraction is evaluated in terms of the following metrics. SIR and SDR are computed using BSS\_EVAL~\cite{bsseval}. The perceptual quality of the extracted sources is quantified using the ``perceptual evaluation of speech quality'' (PESQ~\cite{pesq}) or ``short-time objective intelligibility measure'' (STOI,\cite{stoi}). These metrics are evaluated over the entire signal lengths with the exception of the Dynamic dataset, for which (due to source movements) the measures are evaluated within intervals of length $1$~s each and subsequently averaged. The metrics are either stated as values or as improvements with respect to the mixture (iSIR, iSDR, iPESQ, iSTOI).

When the extraction algorithm fails to track a moving SOI (the SOI moves out of the spatial focus of the method), the desired speech vanishes from the estimated signal. To measure this phenomenon, we also provide \textit{the standard deviation} of the ``SOI Attenuation'' metric, defined as $\sum_k |\hat{s}^k_{\ell}|^2 /$ $\sum_k|s^k_{\ell}|^2$, where $\hat{s}^k_{\ell}$ is the estimate of $s^k_{\ell}$. For a properly extracted moving SOI, this deviation should be close to zero and it increases if the gain of the desired speech fluctuates.

All the experiments have been performed without any adaptation of the algorithm or the FSMN network to a specific scenario. The enrollment set always consists of $1$~minute of speech for each target speaker considered in the given scenario, augmented by reverberation as described in Section~\ref{sec:pilot}. The FSMN pooling context length is $L_c=11$.

\subsection{CSV model for extraction of a moving SOI}
\label{sec:case:base}

This experiment is performed on the Dynamic dataset. It demonstrates the benefits of the CSV-model on mixtures with moving sources and the ability of $\PilotX$ to direct the extraction towards a moving SOI. The deflation is not applied in these experiments, since the mixtures are $25$~s long and $\PilotX$ founds sufficient number of frames to successfully guide the extraction. The results of CSV-AuxIVE are compared to the fully static (FS-IVE, \cite{scheibler2019independent}) and the block-wise static (BS-IVE, \cite{loc:bse:xrob}) variants of AuxIVE. The name of a method followed by subscript $L_T$ (e.g. BS-IVE$_{200}$) denotes the number of frames within the analyzed block.

CSV-AuxIVE and FS-IVE process the entire mixture as a whole using $50$ iterations. BS-IVE processes each block independently and applies $5$ iterations to each block of length $L_T$ and shift $L_T/4$ frames. The inner statistics in BS-IVE are accumulated using recursive forgetting with $\alpha=0.3$ (see \cite{loc:bse:xvec}). All these methods are initialized using the location of the SOI at the beginning of the recording; BS-IVE initializes the extraction at each block by the solution from the previous one. The NFFT length is $1,024$ and shift $200$ samples. The threshold $\thro=2$.

All criteria in Table~\ref{tab:exp:base} indicate that the pilot-guided methods extract the SOI more precisely than the methods relying on initialization (without any pilot). Due to the limited identification accuracy, the performance with $\PilotX$ is inferior to that with $\PilotO$ (by $1.6-2.8$~dB of iSIR). The CSV-AuxIVE achieves superior (or at least comparable) performance compared to its static or block-wise static counterparts. This is notable especially when $\PilotX$ is used. CSV-AuxIVE appears to be more robust than BS-IVE with respect to pilot inaccuracies. The performance of the method decreases with increasing reverberation. However, even when $T_{60}=600$~ms, the CSV-AuxIVE + $\PilotX$ is able to achieve iSIR $6.9$~dB. The iSDR is low in this case, which means that the suppression of interference/noise introduces some distortions into the estimated SOI. However, this scenario is very challenging for spatial filtering due to the low direct to reverberation ratio (the SOI distance is $1.5$~m) and rather high movement speed of the sources. 

The important parameter of CSV-AuxIVE/BS-IVE is the length of block $L_T$, which influences the compromise between adaptivity to movement and the amount of available data. Excessively long blocks (FS-IVE or BS-IVE$_{800}$) yield high iSIR and iPESQ but also increase Attenuation compared to the suitable block length (BS-IVE$_{200}$). Using long blocks, the methods are unable to adapt well to the source movements and the SOI moves out of their spatial focus (the sound vanishes for certain time intervals).\footnote{Note that the Attenuation describes the vanishing of the SOI well for T$_{60}\leq300$~ms but fails to capture this phenomenon for more reverberant scenario. We can observe that this fact is due to the reverberation of the SOI, which is still present in the estimate even when the location (direct path) of the SOI lies outside of the spatial focus of the methods.} The increased Attenuation is observable for the CSV$_{800}$ as well; the increase of iSIR/iPESQ is, however, not present. The prolongation of inner blocks does not bring the advantage of more available data. Application of an insufficiently short block ($50$~frames) allows for good adaptation (low SOI Attenuation), but the overall IS suppression is deteriorating (low iSIR).

\subsection{Properties and limitations of piloting}
\label{sec:exp:pilot}

This Section analyzes the accuracy of the frame-wise speaker identification. Subsequently, the influence of inaccurate pilot on the extraction accuracy is investigated and the causes of pilot failures are discussed.

\subsubsection{The frame-wise dominant speaker identification}
\label{sec:case:xacc}

\begin{figure*}
\begin{minipage}{0.2\textwidth}
\includegraphics[width=\linewidth]{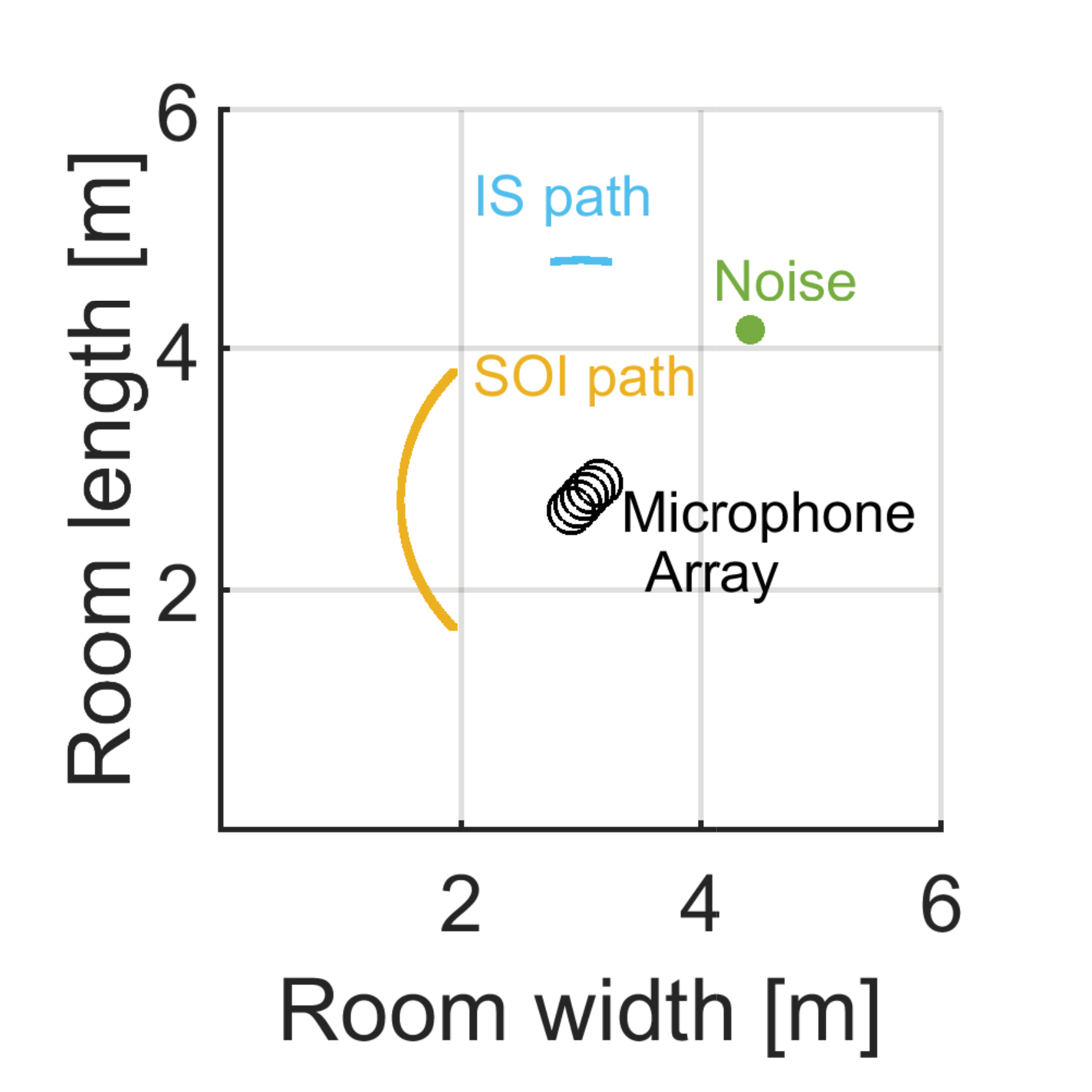}
\caption{\label{fig:base:movs}  Source trajectories and locations for the Dynamic dataset.}
\end{minipage}
\quad
\begin{minipage}{0.38\textwidth}
    \centering
    \includegraphics[width=0.95\linewidth]{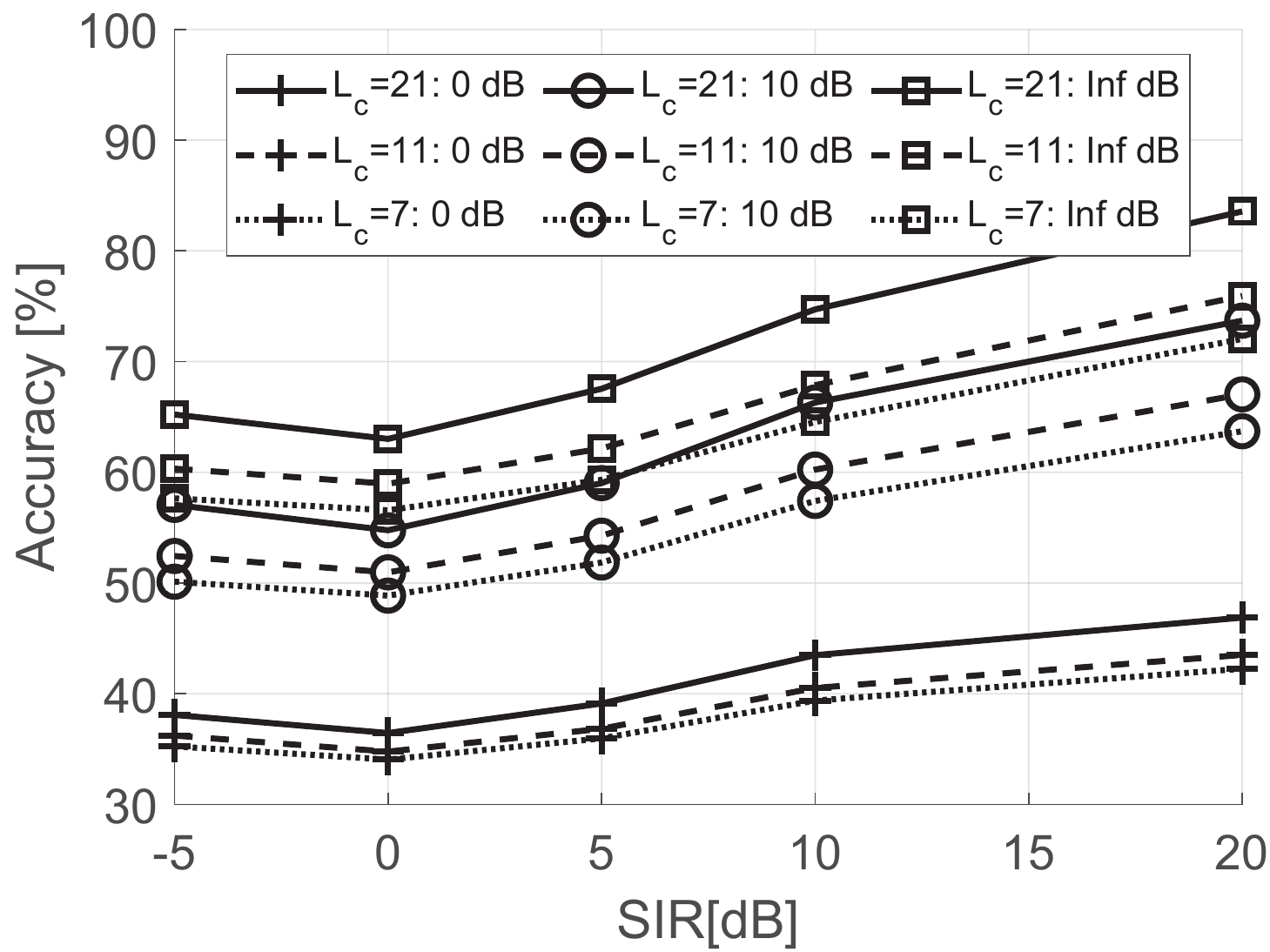}
    \caption{\label{fig:xacc:domspk}  Accuracy in the task of the dominant speaker identification; each marker corresponds to a different SNR.}
\end{minipage}    
\quad
\begin{minipage}{0.38\textwidth}
    \includegraphics[width=0.95\linewidth]{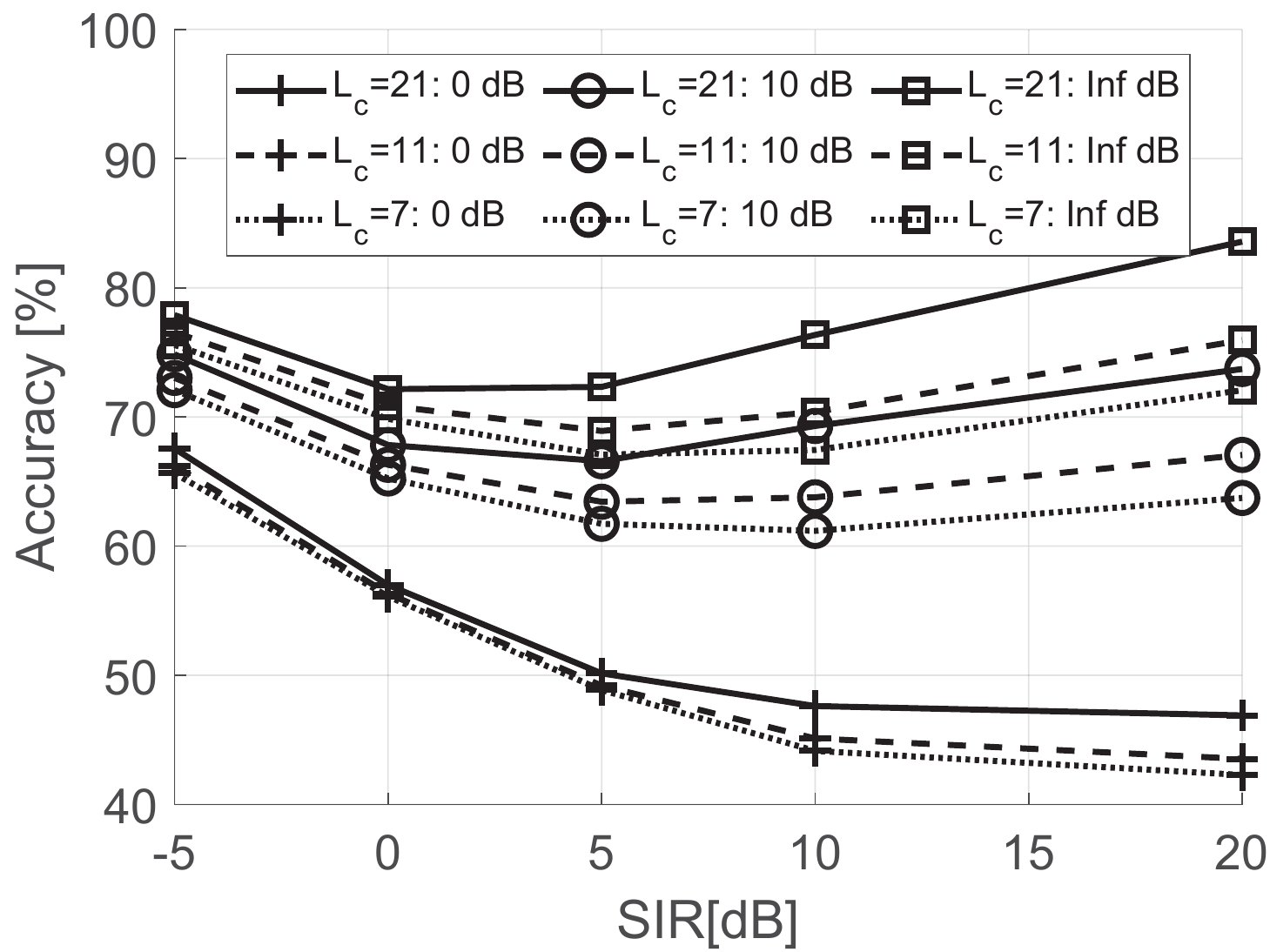}
    \caption{\label{fig:xacc:domsoi} Accuracy in the task of the SOI dominance identification; each marker corresponds to a different SNR.}
\end{minipage}
\end{figure*}

\begin{figure}
\begin{center}
    \centering
    \includegraphics[width=0.72\linewidth]{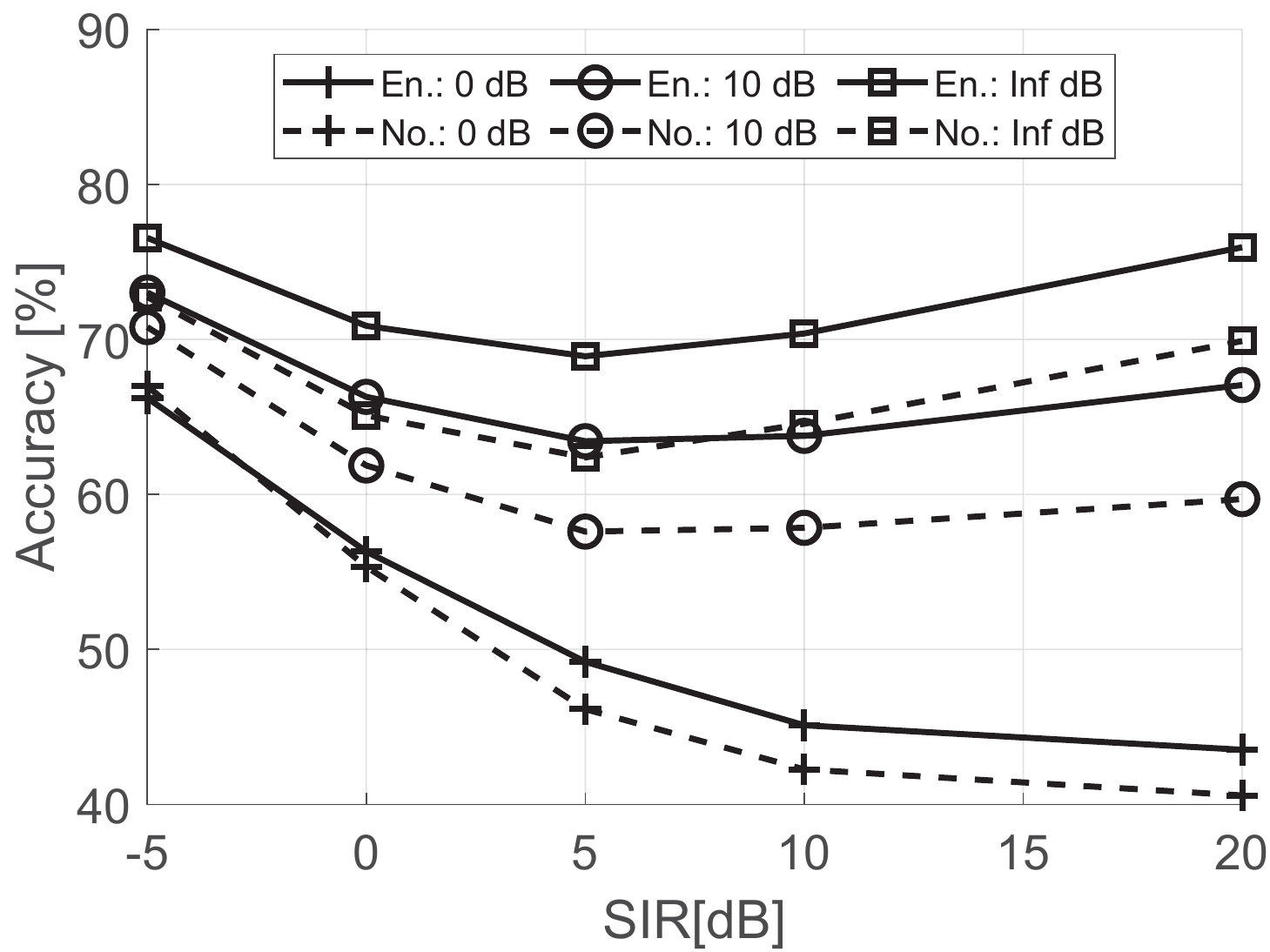}
    \caption{\label{fig:xacc:langdep} Accuracy in the task of the SOI dominance identification with respect to language of speakers in the enrollment set; $L_c=11$ and each marker corresponds to a different SNR.}
\end{center}
\end{figure}

As a ground truth in this task, we use the true identity of the speaker with the highest energy in the mixture. This energy is computed using the same context of frames as the pooling context of FSMN ($L_c \in \{7,11,21\}$, i.e., $\{9,14,26\}$~ms). The accuracy of the identification is thus computed by a comparison of regions determined by the X-vectors and the oracle information obtained using the true energies. The most reverberant part ($T_{60}=600$~ms) of the Dynamic dataset is revisited. Multiple variants of this dataset are considered, each changing the input SIR $\in \{-5,0,5,10,20\}$~dB and the input SNR $\in \{0,10,\infty\}$~dB. Markers in \cref{fig:xacc:domspk,fig:xacc:domsoi} correspond to the averaged accuracy over all mixtures in one such variant. 

Let us first verify the assumption that the source with the highest PLDA score is also the dominant one in the mixture. Considering $L_c=11$ and the noiseless case, Fig.~\ref{fig:xacc:domspk} confirms our assumption with the accuracy ranging from $59\%-77\%$. By definition of the pilot in~\eqref{eq:pilot}, the identity of the interfering speaker is irrelevant for $\PilotX$. The classification is thus simplified to a binary task whether the SOI or an arbitrary other source is dominant. Fig.~\ref{fig:xacc:domsoi} shows that the accuracy of SOI dominance identification is $69\%-77\%$. The presence of noise decreases the accuracy to $63\%-73\%$. The results of the extraction in the previous Section indicate that such accuracy leads to a functional $\PilotX$, which improves the performance of CSV-AuxIVE by iSIR$=2.3$~dB over its non-piloted counterpart. Utilization of $\PilotO$ leads to another increase by $2.8$~dB. The influence of inaccurate pilot on extraction performance is further investigated in Section~\ref{sec:pilot:limits}.

It might seem surprising that accuracy of SOI dominance identification is high despite the low SIR. This is caused by a low occurrence of frames with a dominant SOI (for SIR$=-5$~dB, only $28.5\%$ of frames). The classifier is thus often correct when it assigns the frame to the easily classifiable interfering source with high energy. 

A short context of the pooling layer $L_c$ is required for the frame-wise identification. However, it deteriorates the accuracy due to the increased variability of the X-vectors (less data is available for the pooling). \cref{fig:xacc:domspk,fig:xacc:domsoi} indicate that this accuracy is, as expected, highest for $L_c=21$ and monotonically deteriorates with decreasing $L_c$. On the other hand, Table~\ref{tab:xacc:contlc} shows that the long context $L_c=21$ achieves the worst extraction performance; the piloting is no longer well localized in time. As a compromise, context $L_c=11$ is utilized throughout this manuscript.

\begin{table}[t]
\centering 
\setlength{\tabcolsep}{2.2pt}
\caption{Dynamic dataset: the extraction performance of piloted CSV-AuxIVE ($L_T=200$) with respect to X-vector context $L_c$.}
\label{tab:xacc:contlc}
\begin{tabular}{|c|ccc|ccc|}\hline
{\bf Context}&\multicolumn{3}{c|}{\bf iSDR [dB]}&\multicolumn{3}{c|}{\bf iSIR [dB]}\\ \cline{2-7} 
$L_c$&100ms&300ms&600ms &100ms&300ms&600ms\\ \hline 
7 & 10.25 & 2.24 & -1.83 & 20.48 & 11.40 & 6.71 \\ 
11 & 10.24 & 2.23 & -1.80 & 20.46 & 11.38 & 6.85 \\ 
21 & 10.07 & 2.04 & -1.84 & 20.14 & 10.87 & 6.64 \\\hline 
\end{tabular}
\end{table}

\subsubsection{Language dependence of the SOI identification}
\label{sec:case:langdep}

The blind CSV-AuxIVE algorithm is language independent. However, piloting using $\PilotX$ is based on deep-learning and thus is designed to work on English language present in the training dataset. Its accuracy might deteriorate if applied to an unseen language. To quantify, this scenario compares the SOI identification/extraction achieved on English with results yielded on unseen Norwegian. It analyzes a slightly modified version of the Dynamic dataset. The original English speakers are replaced by four Norwegian ($2$ male and $2$ female) originating in the NST speech database~\cite{data:nst}. 

The results in Fig.~\ref{fig:xacc:langdep} corroborate that X-vectors are slightly language dependent; the accuracy for Norwegian speakers is lower by about $4.5\%$. However, this does not influence the extraction performance much. The metrics in Table~\ref{tab:exp:langdep} indicate that the non-piloted extraction is slightly less accurate for the Norwegian dataset. This decrease does not stem from the language as such but it is caused by longer silences between Norwegian sentences. When the SOI is quiet, the non-piloted extractor tends to converge to an arbitrary active source. The utilization of a pilot completely removes this difference. The results for CSV-AuxIVE piloted via $\PilotX$ are comparable for both datasets (difference is maximally $1$~dB in iSIR and iSDR); the proposed method can thus be considered language independent in this experiment.

\begin{table}[t]
\centering 
\setlength{\tabcolsep}{2.2pt}
\caption{Dynamic dataset: the extraction performance with respect to spoken language (English or Norwegian); the language dependence of $\PilotX$. The subscript ${L_T}$ denotes the number of frames within the analyzed block.}
\label{tab:exp:langdep}
\begin{tabular}{|l|l|c|cc|cc|cc|}\hline
\multirow{2}{*}{\bf Method} & \multirow{2}{*}{\bf Pilot}& \multirow{2}{*}{\bf Lang.}&\multicolumn{2}{c|}{\bf iPESQ}&\multicolumn{2}{c|}{\bf iSDR [dB]}&\multicolumn{2}{c|}{\bf iSIR [dB]}\\ \cline{4-9} 
 &  &  &100ms&600ms &100ms&600ms &100ms&600ms \\ \hline 
$\text{CSV}_{200}$ &  \multicolumn{1}{c|}{-} & Eng. & 0.49 & 0.02 & 7.89 & -2.88 & 17.03 & 4.60\\ 
$\text{CSV}_{200}$ &  $\PilotO$ & Eng. & 0.87 & 0.13 & 11.52 & 0.16 & 22.01 & 9.64\\ 
$\text{CSV}_{200}$ &  $\PilotX$ & Eng. & 0.76 & 0.06 & 10.24 & -1.80 & 20.46 & 6.85\\\hline 
$\text{CSV}_{200}$ &  \multicolumn{1}{c|}{-} & Nor. & 0.42 & -0.01 & 7.37 & -3.12 & 15.05 & 2.62\\ 
$\text{CSV}_{200}$ &  $\PilotO$ & Nor. & 0.81 & 0.11 & 12.54 & 0.80 & 22.42 & 9.70\\ 
$\text{CSV}_{200}$ &  $\PilotX$ & Nor. & 0.74 & 0.02 & 11.14 & -1.52 & 20.86 & 6.02 \\\hline 
\end{tabular}
\end{table}

\subsubsection{Limitations of the embedding-based piloting in low SIR scenarios}
\label{sec:pilot:limits}

By definition in~\eqref{eq:pilot}, the pilot is non-zero/active when the SOI is dominant in a subset of frames. This condition becomes difficult to fulfill when SIR is low. Let us demonstrate using mixtures in the Dynamic dataset. Considering three levels of SIR=$\{20,0,-5\}$~dB; the SOI is dominant in $93.2\%, 48.3\% \text{ and } 28.5\%$ of frames, respectively. For a low SIR, the potential support is limited, which weakens the guidance provided by the pilot. $\PilotX$ suffers from a further reduction of the support, because it incorrectly identifies a subset of the dominant frames. An extreme case of the pilot being equal to zero for all frames leads to non-piloted extraction (which, moreover, tends to extract the dominant interfering source).

The deflation approach provides a mechanism to alleviate these limitations. Let us demonstrate via an extraction experiment on MC-WSJ0-2mix dataset~\cite{mc-wsj0-2mix}. The dataset contains $3,000$ mixtures of two active speakers. Since each speaker can assume the role of the SOI, $6,000$ independent extractions can be performed. Let us observe in Table~\ref{tab:pil:fails} the number of cases when CSV-AuxIVE successfully extracts a source, but it is an unwanted source due to insufficient guidance. We assume this happens when the iSDR is less than $-2$~dB. The non-piloted CSV-AuxIVE fails in $2,986$ cases. It has no way to focus on a specific SOI and, in addition, fails to process some of the mixtures (output SDR is close to $0$~dB). $\PilotX$ reduces the fail rate by about $77$~\% to $680$ cases; in $440$ of these mixtures, the SOI is the weaker source (input SDR$<0$~dB). The deflation significantly reduces the number to $58$ cases, which is comparable to the utilization of $\PilotO$ (which does not suffer from the erroneous classification of frames).

\begin{figure}
\centering
\includegraphics[width=0.49\linewidth]{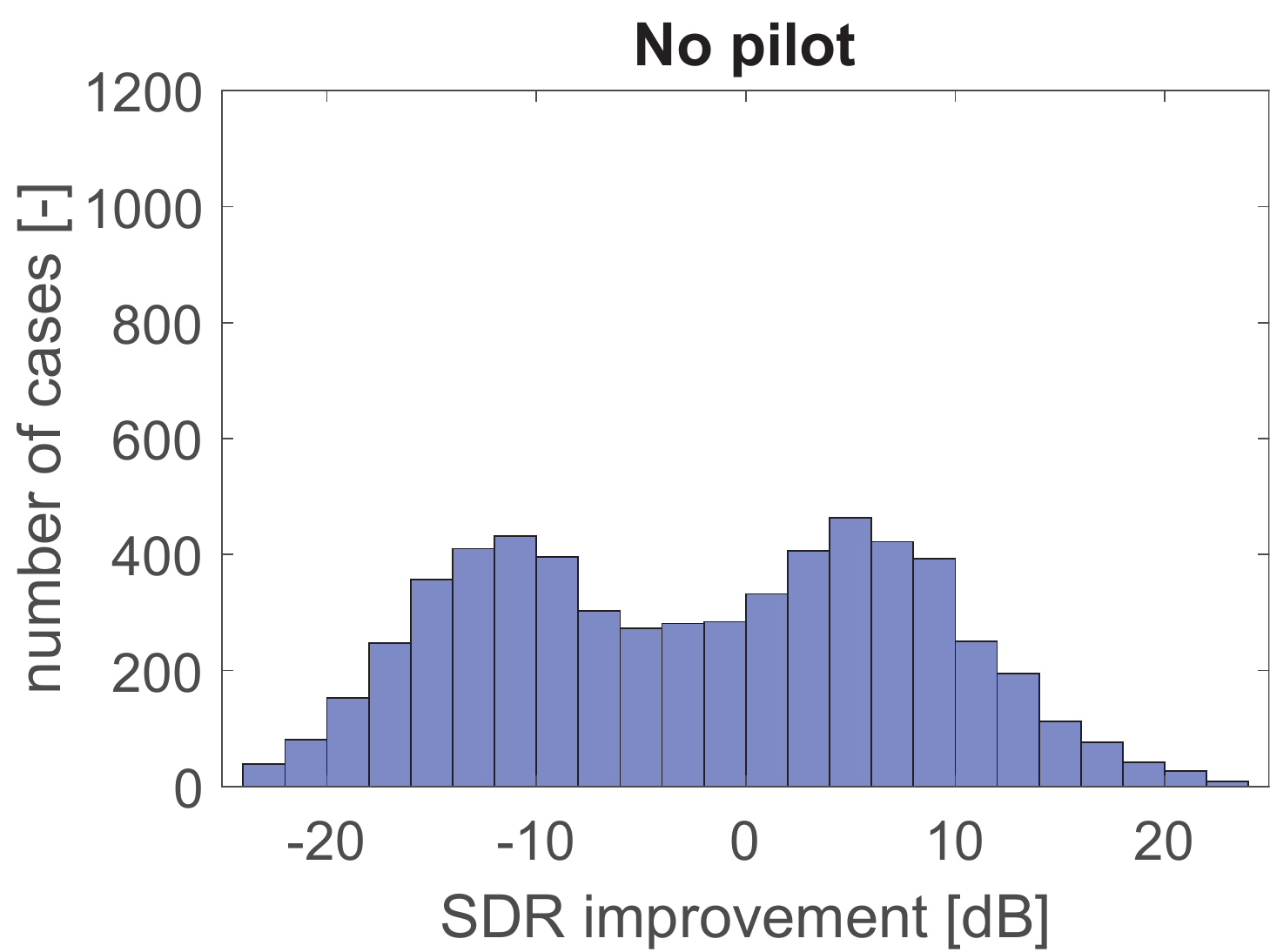}
\includegraphics[width=0.49\linewidth]{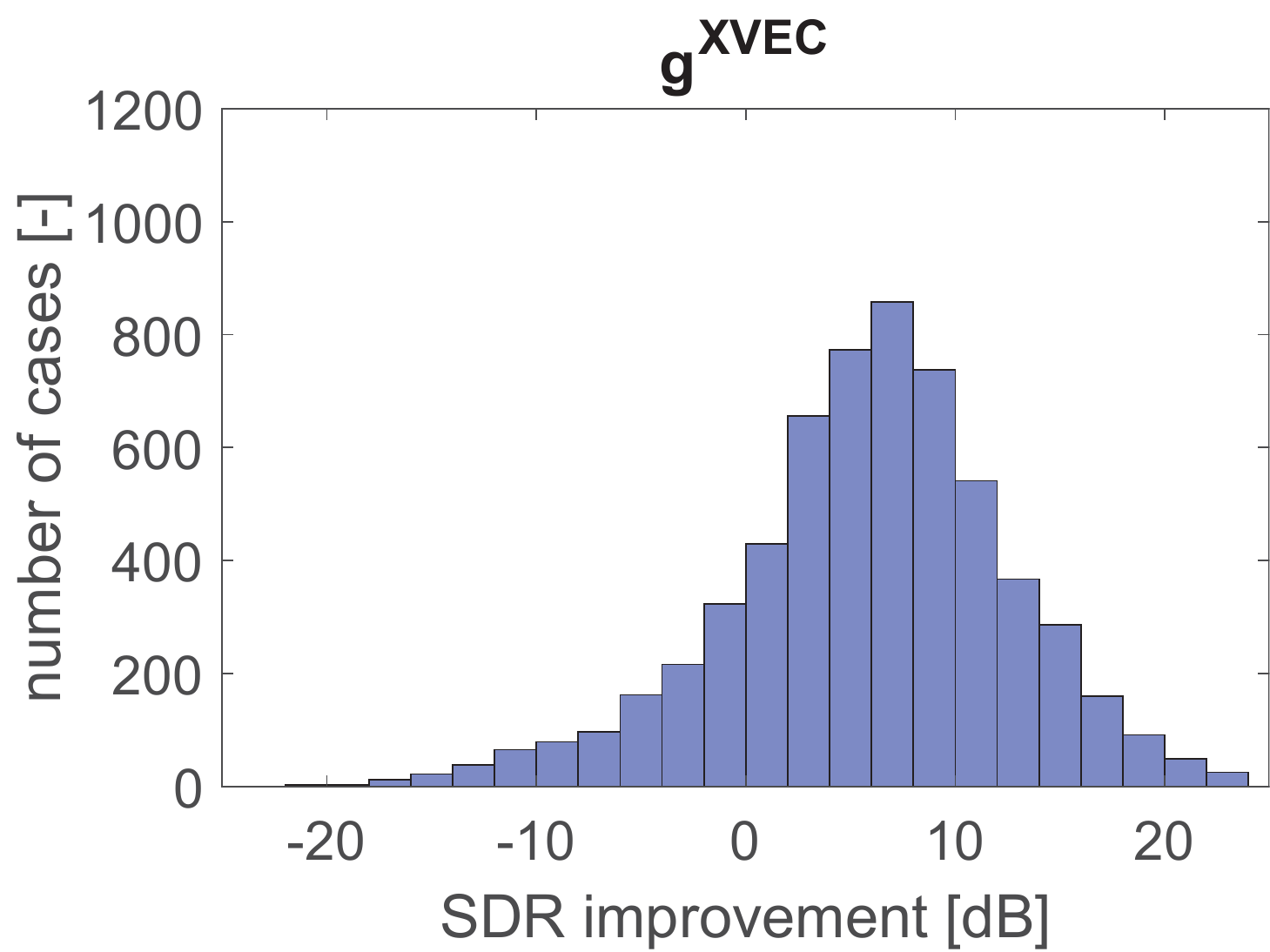}
\includegraphics[width=0.49\linewidth]{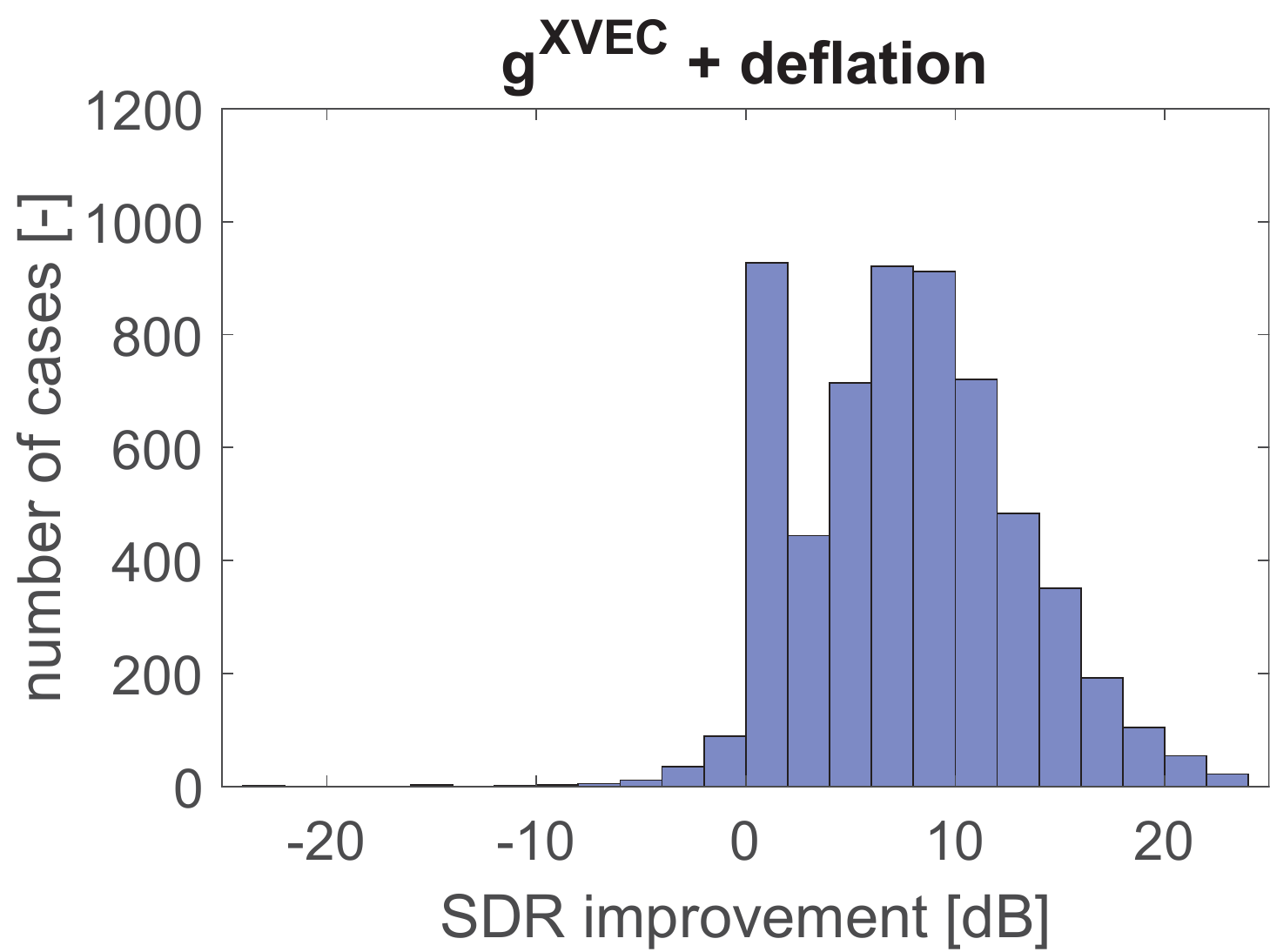}
\includegraphics[width=0.49\linewidth]{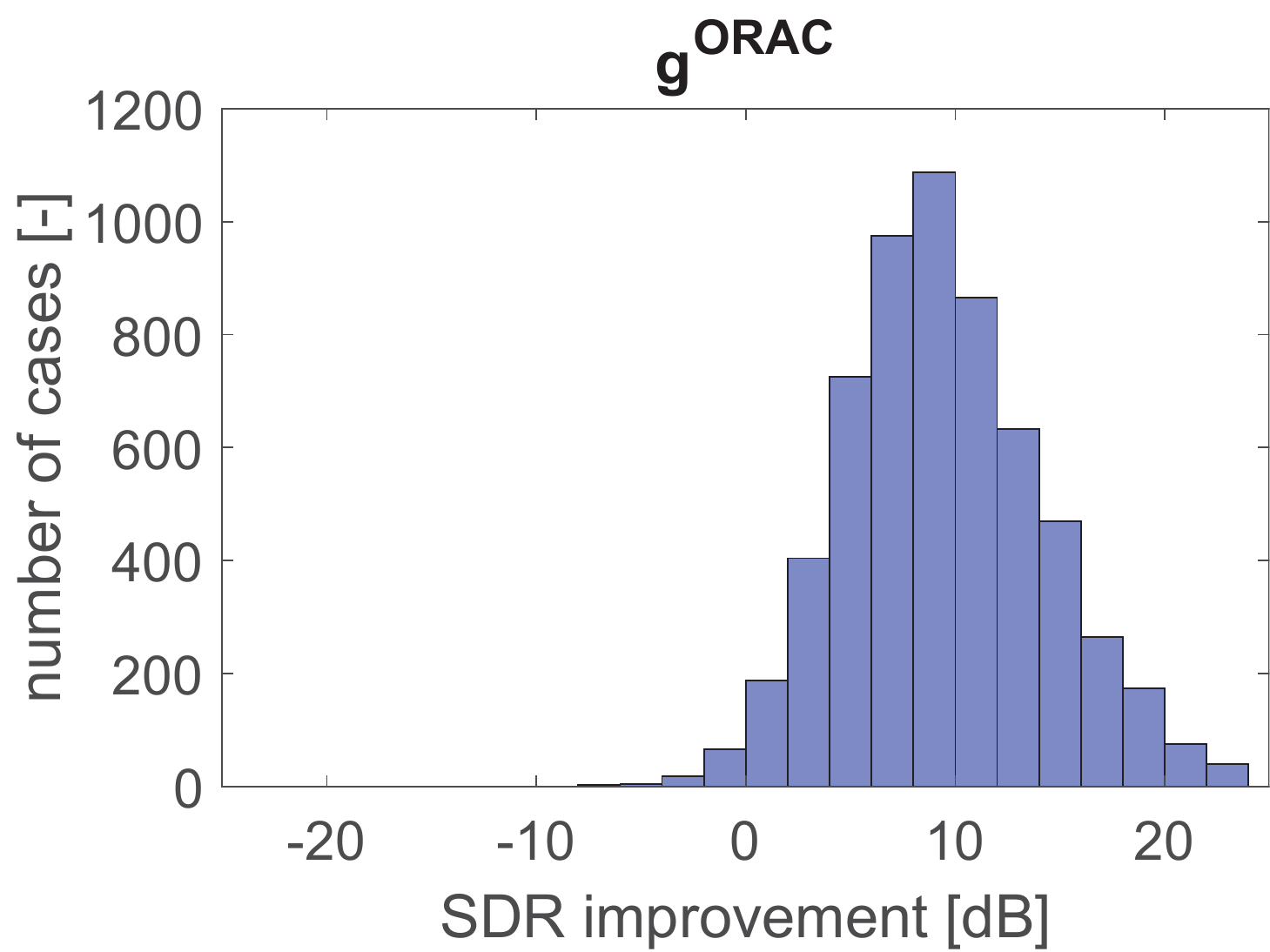}
\caption{\label{fig:exp:pifails:hist} MC-WSJ0-2mix, $4$~channels: iSDR distributions achieved by CSV-AuxIVE endowed with various forms of guidance.}
\end{figure}

The distributions of iSDR achieved in this task are shown in Fig.~\ref{fig:exp:pifails:hist}. The CSV-AuxIVE without a pilot has a symmetric distribution of iSDR because it cannot focus on specific SOI. Utilization of $\PilotX$ shifts the distribution towards the positive iSDR. However, many cases of negative iSDR remain, corresponding to unsuccessful piloting. For some cases, the piloting prevents extraction of an unwanted source, but fails to guide the extraction towards the SOI. Therefore, CSV-AuxIVE+$\PilotX$ yields a slightly increased number of cases with no extracted source compared to CSV-AuxIVE without a pilot (see Table~\ref{tab:pil:fails}). The deflation manages to remedy almost all failed piloting cases and further shifts the distribution to the positive values. However, part of these remedied cases does not lead to successful extraction of the SOI; their output iSDR is equal to $0$~dB. This effect is caused by an overly conservative behavior in the assessment of the extraction quality (see Section~\ref{sec:exp:asses} for further discussion). It recognizes that an unwanted source was extracted and performs the deflation of the mixture. However, the reduced mixture is not recognized as a better estimate of the SOI than the original mixture. Consequently, the original mixture is returned as the SOI estimate. The utilization of an accurate $\PilotO$ causes a successful extraction of the SOI for most of the mixtures.

\begin{figure}
\centering
\includegraphics[width=0.72\linewidth]{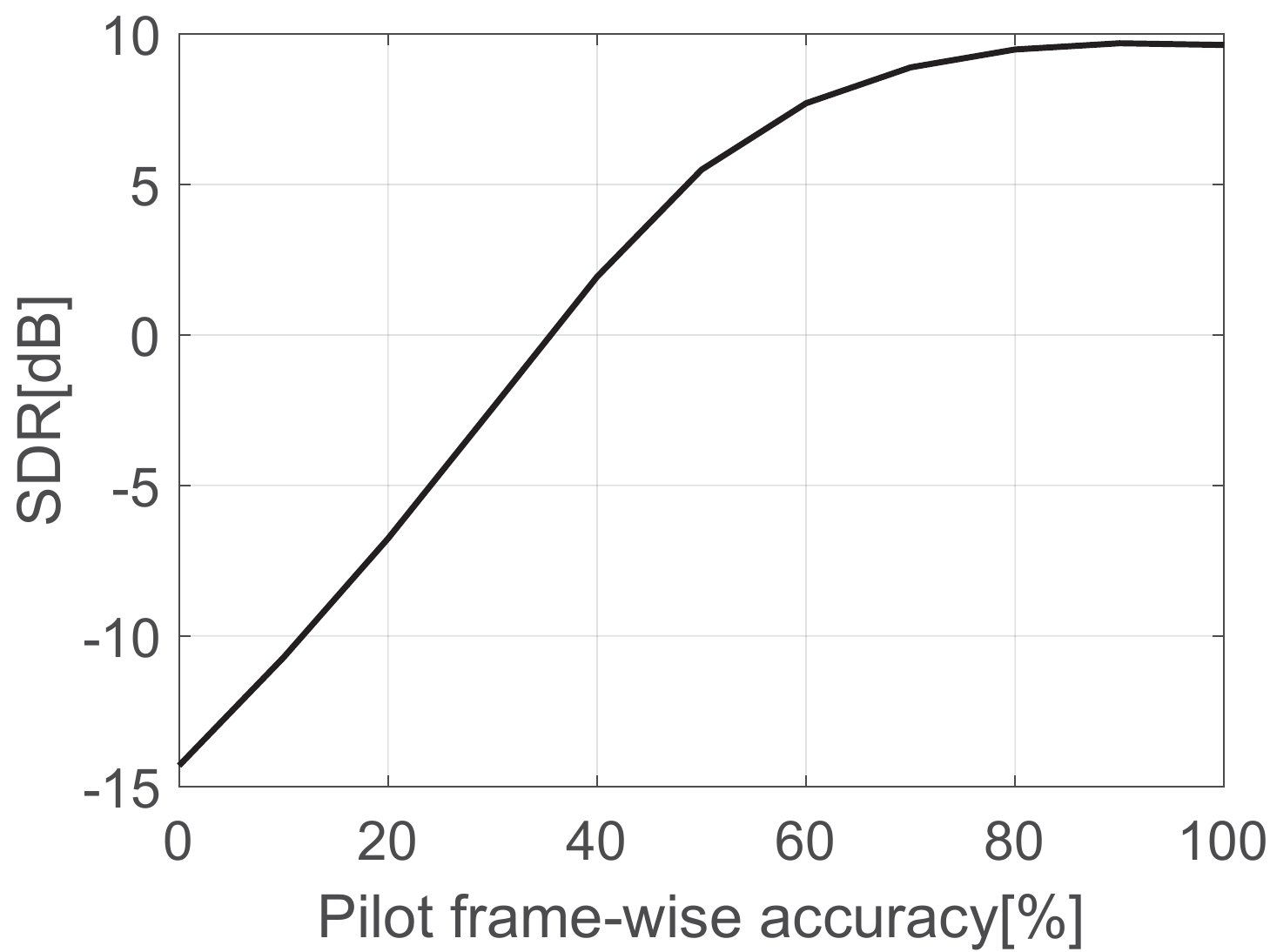}
\caption{\label{fig:exp:pilot:acc2sdr} MC-WSJ0-2mix, $4$~channels: SDR achieved by the CSV-AuxIVE using oracle pilot, whose frames corresponding to SOI are gradually interchanged with frames dominated by the interfering source.}
\end{figure}

The influence of the incorrectly classified frames in the pilot on the final SDR is shown in Fig.~\ref{fig:exp:pilot:acc2sdr}. In this experiment, we pilot the extraction on the MC-WSJ0-2mix dataset by $\PilotO$. We gradually replace $10\%$ of frames with dominant SOI by $10\%$ of frames corresponding to the unwanted source. The frames with comparable energy are swapped first; the frames with a highly dominant source are swapped as the last ones. It can be seen that the substitution of about $20\%$ of frames does not significantly influence the performance. When all frames are substituted, i.e., the pilot contains only frames corresponding to the interfering source, CSV-AuxIVE achieves highly negative SDR because it is in all cases guided towards the interfering source.

The accuracy of SOI dominance identification in $\PilotX$ is $66.3\%$ on the MC-WSJ0-2mix dataset, which yields an SDR of $6$~dB. Comparing the results with Fig.~\ref{fig:exp:pilot:acc2sdr}, such accuracy should yield an SDR of about $8$~dB. The modeling of errors by distorting the $\PilotO$ thus appears to be slightly more optimistic than the results achieved using the realizable $\PilotX$.

\begin{table}
\centering
\caption{MC-WSJ0-2mix, $4$~channels: the number of cases, when CSV-AuxIVE: 1) extracts an unwanted source due to insufficient piloting (iSDR < $-2$~dB), 2) extracts no source ($2$~dB < iSDR < $-2$~dB), 3) successfully extracts the SOI (iSDR > $2$~dB).}
\label{tab:pil:fails}
\begin{tabular}{|c|c|c|c|} \hline
 \textbf{Pilot/deflation} & \textbf{Unwanted} & \textbf{No source}  & \textbf{SOI} \\
  & \textbf{source} & \textbf{extracted}  & \textbf{extracted} \\
  & \textbf{extracted} &  &  \\
  \hline \hline 
 No pilot & 2986 & 616 & 2398 \\ \hline
 $\PilotX$ & 697 & 753 & 4550\\ \hline
 $\PilotX$ + deflation & 58 & 1016 & 4926 \\ \hline
 $\PilotO$ & 24 & 253 & 5723 \\ \hline
\end{tabular}
\end{table}

\subsection{Non-intrusive assessment of extraction quality}
\label{sec:exp:asses}

This section verifies whether the PLDA score can be used to select a superior SOI estimate within several available variants. The superiority is measured using the standard objective and perceptual metrics (SIR, SDR, PESQ, STOI).

We use two datasets: the simulated development part of the CHiME-4 dataset contains $1,640$ mixtures of speech (produced by $4$~speakers) and noise, whereas MC-WSJ0-2mix contains $3000$ mixtures of two utterances (produced by $18$ speakers). The non-piloted CSV-AuxIVE with uniform initialization is applied to these recordings and stopped consecutively after $\{0,5,10,15,20,25\}$ iterations for the CHiME-4 and $\{0,15,30,50\}$ iterations for MC-WSJ0-2mix. For each utterance and each stop, the PLDA score $M(\enrS,\hrS)$ and the metrics are evaluated. Subsequently, the differences with respect to the previous stop are computed because the goal is to find the relationship between the change of $M(\enrS,\hrS)$ and the change in the metrics.

\begin{figure}
\centering
\includegraphics[width=0.49\linewidth]{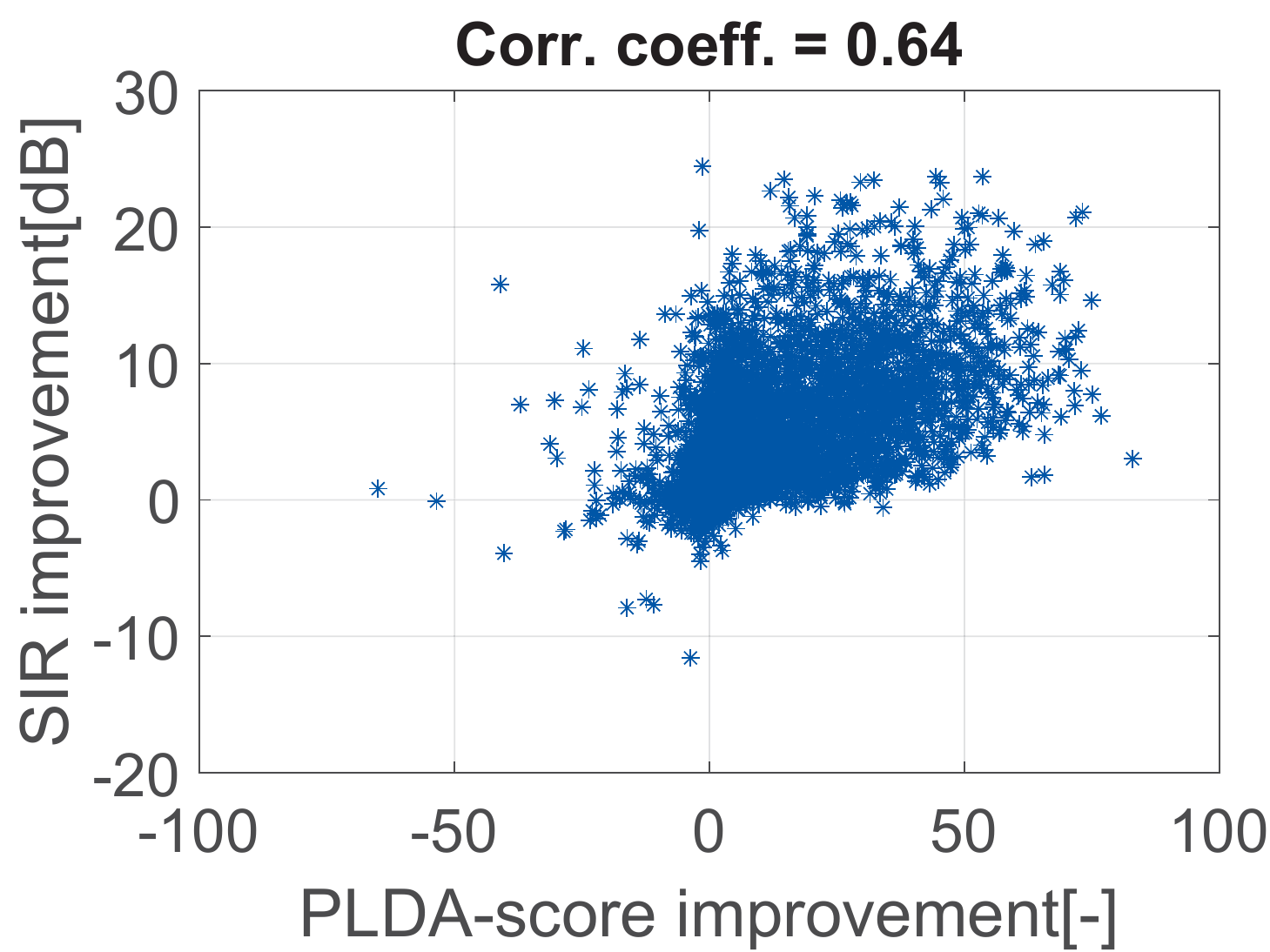}
\includegraphics[width=0.49\linewidth]{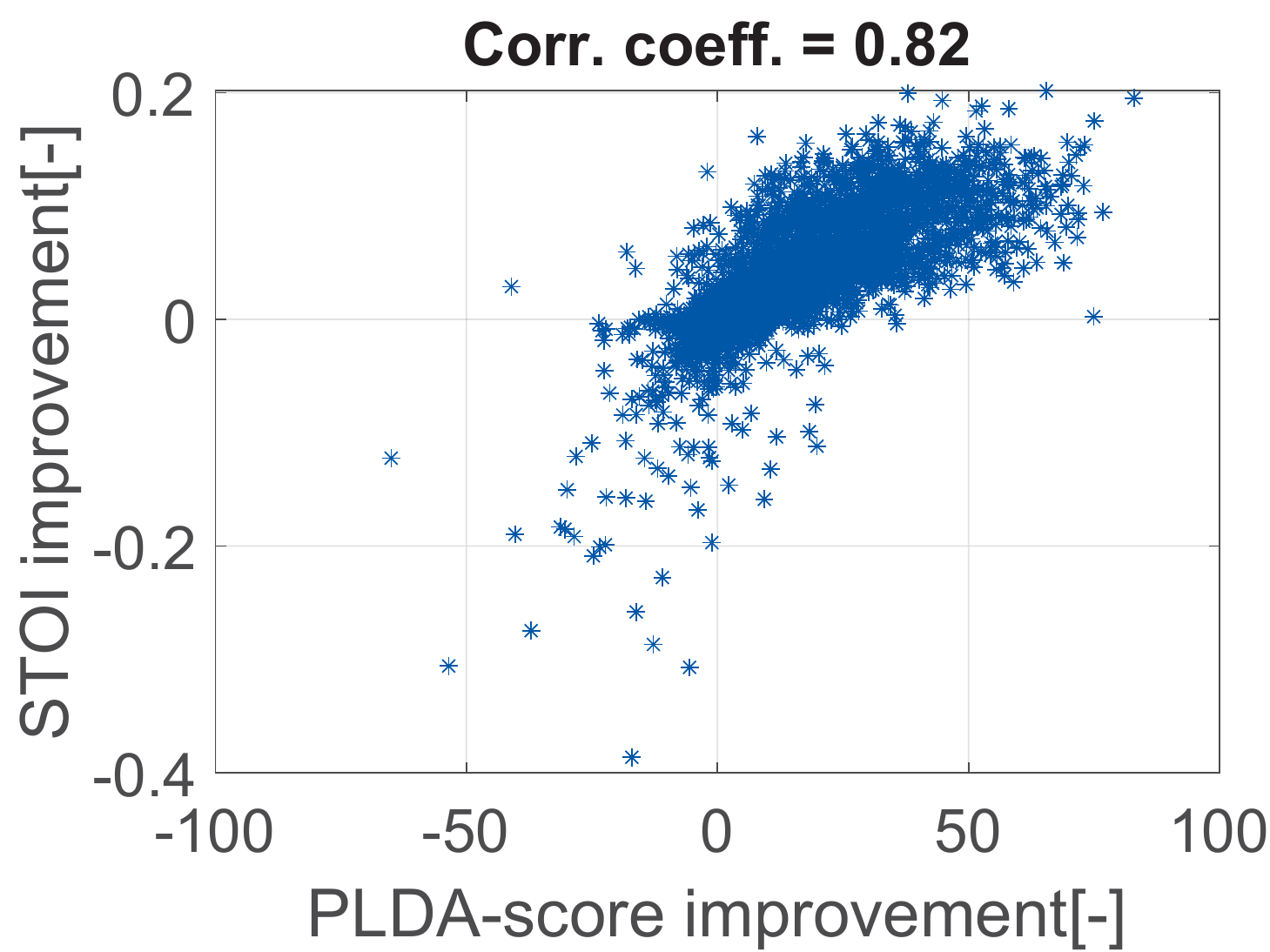}
\caption{\label{fig:exp:qual2} CHiME-4 (simulated development part): dependency between the improvements of the objective criteria and the improvements of PLDA score on speech-noise mixtures.}
\vspace{2mm}
\includegraphics[width=0.49\linewidth]{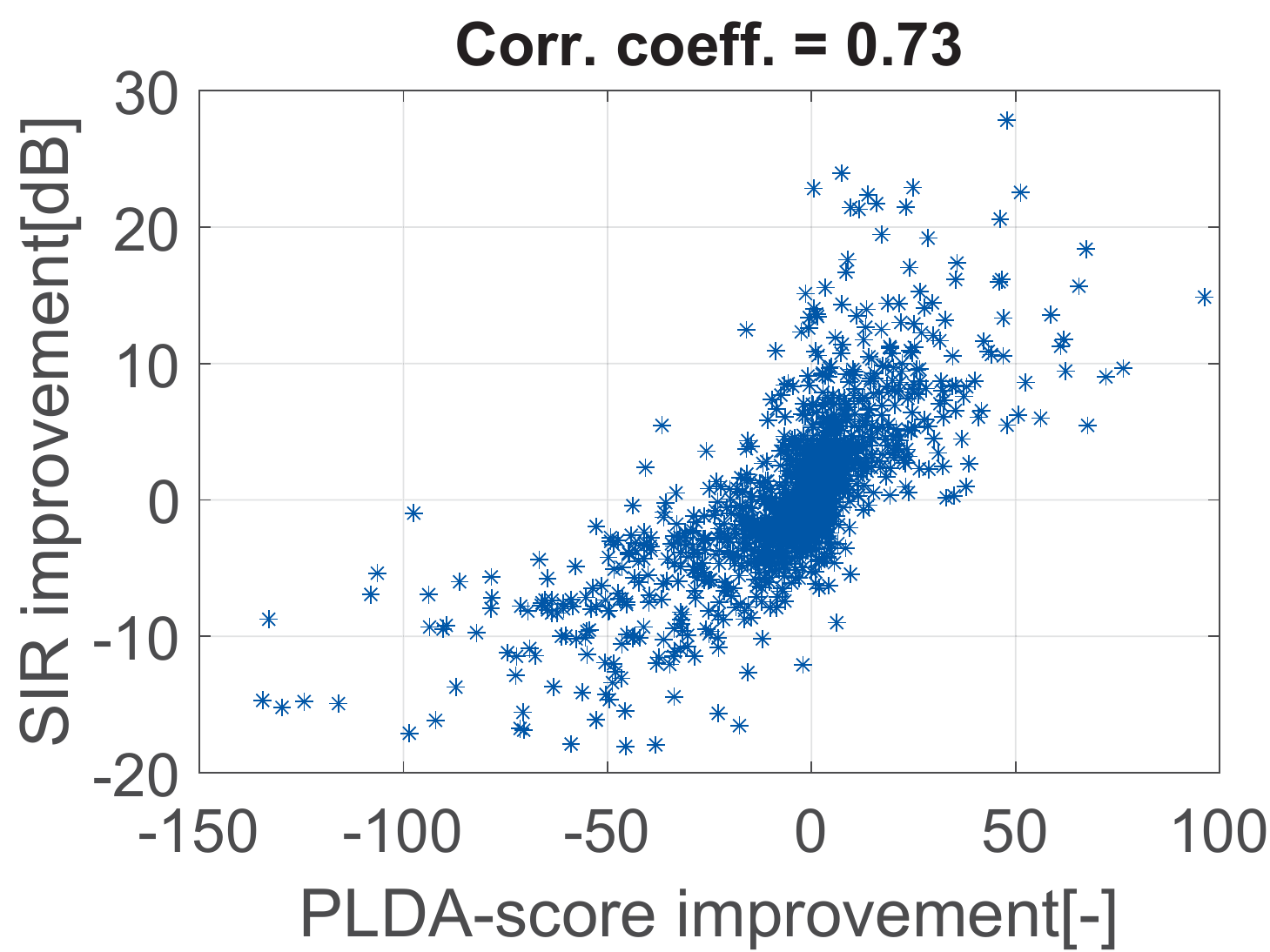}
\includegraphics[width=0.49\linewidth]{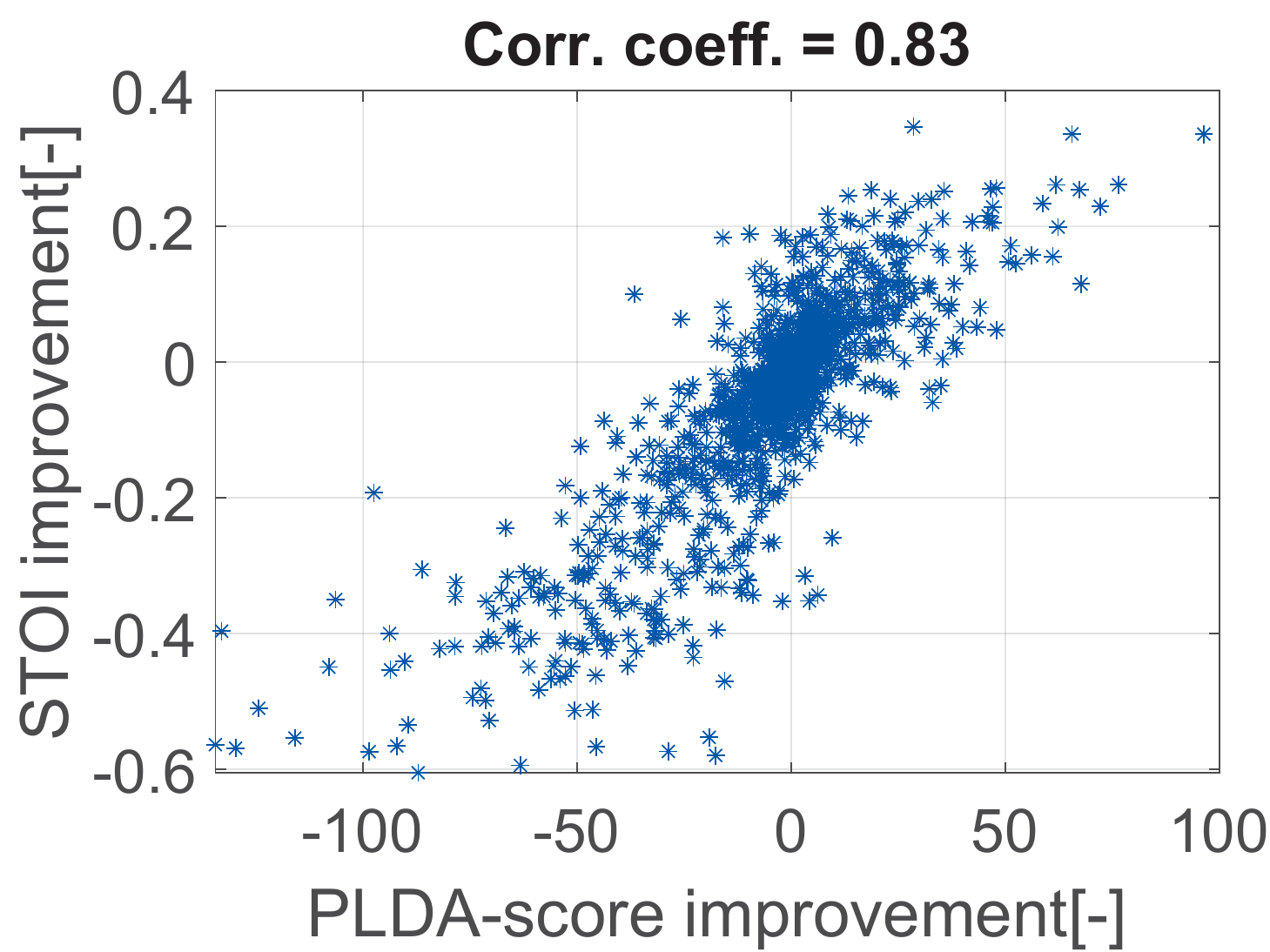}
\caption{\label{fig:exp:qual:wsj0} MC-WSJ0-2mix: dependency between the improvements of the objective criteria and the improvements of PLDA score on speech-speech mixtures.}
\end{figure}

The differences plotted in Figs.~\ref{fig:exp:qual2} and~\ref{fig:exp:qual:wsj0} indicate the existence of a linear dependence. The Pearson correlation coefficient reaches a value of $0.83$ for STOI. From another perspective, the assessment can also be perceived as a binary classifier: given the increase/decrease of $M(\enrS,\hrS)$, we want to predict the respective change in the objective criterion. Tables~\ref{tab:exp:qual2} and~\ref{tab:exp:qual:wsj0} show that the classification accuracy is $72.7$~\% and $75.5$~\% for SIR on speech-noise and speech-speech mixtures, respectively. 

There are two types of error: 1) False positives ($M(\enrS,\hrS)$ increases, but the metrics decrease) are more severe and potentially lead us to select an interfering source. Fortunately, the number of cases with significant deterioration is not very high. A decrease worse than $1$dB in SIR happens only in $5.5$\% of cases for speech-speech mixtures. 2) False negatives ($M(\enrS,\hrS)$ decreases, but criteria increase) potentially lead us to a selection of an inferior estimate. An $8.4$\% proportion of the cases exhibits a significant decrease in SIR for speech-speech mixtures. These cases cause the overly conservative behavior of the deflation described in Section~\ref{sec:pilot:limits}. 

The proposed assessment is functional in both speech-noise and speech-speech mixtures. However, the number of significant incorrect cases is larger for the speech-speech mixtures, where the active sources are more similar and can be confused more easily.

\begin{table}[t]
\centering
\caption{\label{tab:exp:qual2} CHiME-4 (simulated development part, speech-noise mixtures): the evaluation of the proposed extraction assessment serving as a binary classifier of the speech-quality metrics. Significant cases denote samples in which the erroneous increase/decrease in SIR/SDR is larger than $1$dB, $0.01$ in STOI or $0.05$ in PESQ.}
\begin{tabular}{|l|c|c|c|c|}
\hline
                             & \textbf{SIR} & \textbf{SDR} & \textbf{PESQ} & \textbf{STOI} \\ \hline\hline
Correlation coefficient [-]  &  $0.64$  & $0.74$ & $0.58$ & $0.82$    \\ \hline\hline
Accuracy[\%]                 &  $72.7$ & $70.4$ & $70.4$ & $69.2$     \\ \hline
False positives [\%]         &  $5.7$  & $15.3$ & $16.3$  & $20.2$    \\ \hline
Significant false positives [\%] & $0.3$& $1.2$ & $0.8$ & $1.6$     \\ \hline
False negatives [\%]         &  $21.6$ & $14.2$ & $13.3$ & $10.5$    \\ \hline
Significant false negatives [\%] & $7.9$& $2.7$ & $1.7$ & $0.8$     \\ \hline
\end{tabular}
\vspace{2mm}
\caption{\label{tab:exp:qual:wsj0} MC-WSJ0-2mix (speech-speech mixtures): the evaluation of the proposed extraction assessment serving as a binary classifier of the speech-quality metrics. Significant cases denote samples in which the erroneous increase/decrease in SIR/SDR is larger than $1$dB, $0.01$ in STOI or $0.05$ in PESQ.}
\begin{tabular}{|l|c|c|c|c|}
\hline
                             & \textbf{SIR} & \textbf{SDR} & \textbf{PESQ} & \textbf{STOI} \\ \hline\hline
Correlation coefficient [-]  &  $0.73$  & $0.77$ & $0.63$ & $0.83$    \\ \hline\hline
Accuracy[\%]                 &  $75.5$ & $75.6$ & $71.7$ & $76.5$     \\ \hline
False positives [\%]         &  $10.3$  & $12.5$ & $12.5$  & $14.3$    \\ \hline
Significant false positives [\%] & $5.5$& $6.4$ & $5.8$ & $11.0$     \\ \hline
False negatives [\%]         &  $14.3$ & $11.9$ & $15.8$ & $9.1$    \\ \hline
Significant false negatives [\%] & $8.4$& $4.9$ & $7.9$ & $6.1$     \\ \hline
\end{tabular}
\end{table}

\subsection{Extraction via deflation on public datasets}
\label{sec:exp:defl}

The following experiments provide comparison between results achieved by the proposed method and the results reported in the literature. The experiments also show benefits brought by deflation.  

\subsubsection{Extraction of the SOI from noisy recordings with transients and microphone failures}
\label{sec:exp:qual:chime}

Piloting and deflation should not be necessary on CHiME-4 data since the recordings contain only one active speaker. However, the real-world part of CHiME-4 is sometimes distorted by transients and microphone failures. These signals behave like sources that are strongly non-Gaussian, which have wide areas of attraction in contrast functions of blind algorithms. Therefore, they can be extracted instead of speech. The piloting and deflation used in our method provide effective solutions for this phenomenon.

The enhancement via piloted CSV-AuxIVE is compared with two enhancers known to be very successful on the CHIME-4 data: BeamformIt~\cite{beamformit}, a weighted delay-and-sum beamformer, which is used as a front-end algorithm in the original CHiME-4 baseline system. The Generalized Eigenvalue Beamformer (GEV) is a front-end solution proposed in \cite{heymann, heymann_chime4}. The latter represents one of the most successful enhancers for CHiME-4. It relies on voice activity detection (VAD) via deep networks trained specifically for the CHiME-4 data. We utilize the feed-forward topology of the VAD (the training procedure was kindly provided to us by the authors of \cite{heymann}) and re-train the network using the training part of the CHiME-4 data. 

Since the true references of the sources are not available for the real-world part of CHiME-4, the experiments are evaluated using the WER of the original baseline recognizer from \cite{chime_data_soft}. All of the proposed methods are initialized by the relative transfer function estimator from \cite{gannot2001}. CSV-AuxIVE performs $5$~iterations in the STFT domain with an FFT length of $512$, hop-size of $200$ (the shift of the FSMN network) and applied Hamming window; the sampling frequency is $16$~kHz. The length of the CSV-AuxIVE block is $2$~seconds ($L_T=160$ frames). The enrollment set for piloting contains $8$~speakers; respective speech signals originate from the simulated development part of CHiME-4.

\begin{table}
\centering
\caption{\label{tab:exp:chime} WER [\%] yielded on the real-world part of the CHiME-4 datasets. Mixture results are achieved using data from channel $5$.}
\begin{tabular}{|l||c||c|c||c|c|c|}
\hline
& \begin{tabular}[c]{@{}c@{}}\textbf{Mix.}\\ {\bf ch.}{$\bm 5.$}\end{tabular}
& \begin{tabular}[c]{@{}c@{}}\textbf{Beam-}\\ \textbf{form-}\\\textbf{It}\end{tabular} 
& \textbf{GEV} 
& \textbf{CSV}
& \begin{tabular}[c]{@{}c@{}}\textbf{CSV}\\ +pilot\end{tabular} 
& \begin{tabular}[c]{@{}c@{}}\textbf{CSV}\\ +pilot\\ +defl.\end{tabular} \\ \hline
\textbf{Dev.} & $9.8$ & $5.8$ & $4.6$ & $5.8$ & $5.4$ & $5.4$ \\ \hline
\textbf{Test} & $19.9$ & $11.5$ & $8.1$ & $9.9$ & $9.5$ & $9.3$ \\ \hline
\end{tabular}
\end{table}

The results in Table~\ref{tab:exp:chime} indicate that the WER of CSV-AuxIVE\footnote{Slightly different WER of CSV-AuxIVE was reported in~\cite{loc:auxive}; it is caused by a different FFT frame-shift and the number of performed iterations.} is lowered by using piloting and further using deflation. This is in agreement with the discussion presented in Section~\ref{sec:pilot:limits}: namely, the piloting significantly reduces the number of diverged cases and the deflation allows for re-estimation of the SOI when the piloting fails. The proposed method yields lower WER values compared to BeamformIt but is still outperformed by GEV. Nevertheless, GEV is a technique specifically tailored to CHiME-4 due to dataset-specific VAD and is limited to enhancement of recordings without cross-talk. In contrast, the proposed technique is, without adaptation, applicable to both speech enhancement and extraction. Even without piloting, CSV-AuxIVE achieves results approaching those of GEV without a need for training.

\subsubsection{Extraction of the SOI from cross-talk in a reverberant environment}
\label{sec:exp:mcwsj2}

The following experiment compares the performance of the proposed method on the MC-WSJ0-2mix dataset to the results reported in the literature. The competing methods can be divided into three groups: 1) Oracle methods representing ideal extractors. These methods cannot be used in practice as they utilize information that is normally not available. 2) Methods based on machine learning (ML), which rely on the existence of a scenario-specific training dataset. 3) Blind source separation/extraction methods, which exploit spatial information extracted from the multi-channel mixture.

For ML-based methods, we consider extraction approaches that identify the SOI and solely recover this source from the mixture. For blind approaches, the literature usually presents methods performing the complete separation (BSS). Here, all sources in the mixture are estimated (the number of sources must be known), and the SOI is subsequently identified among them. This can be done either in an oracle manner using the true reference during evaluation or using ML-based speaker identification (to this end, we use the same FSMN network as for piloting; the X-vector pooling context $L_c=L$). In contrast, the piloted CSV-AuxIVE extracts (BSE) only the SOI and does not require the number of interfering sources.

The oracle approaches are represented by the 1) multi-channel Wiener filter (MCWF), which uses the oracle covariance matrix of the target speech and constitutes the upper boundary for the extraction based on spatial filtering. The machine learning-based separation is represented by: 2) TasNet from~\cite{tasnet}, which is based on a convolutional topology performing full separation in the time domain; subsequently, the SOI is selected via speaker identification. 3) The frequency (FD) and time domain (TD) variants of SpeakerBeam~\cite{speakerbeam:timedomain,mlse:spkbeamext1:icassp2021}, which perform speaker extraction based on an enrollment utterance. 
Blind methods are represented by 4) masking-based binaural MESSL~\cite{mandel2010}, 5) binaural GCC-NMF~\cite{bss:nmf} based on non-negative matrix factorization, 6) consistent ILRMA from~\cite{bss:ilrma:advsig2020}, 7) GLOSS~\cite{bss:smplx:taslp2020} using sparsity-based spectral masking and single-channel post-filter and 8) static auxiliary function-based independent vector extraction FS-IVE~\cite{scheibler2019independent}.

\begin{table}
\centering
\caption{\label{tab:exp:wsj0} MC-WSJ0-2mix: SDR [dB] yielded using machine-learning (ML), blind source separation (BSS) and blind source extraction (BSE). The column ``Tr. data'' quantifies the volume of the required scenario-specific training data.}
\begin{tabular}{|c|c|c|c|c|c|c|}
\hline
\textbf{Approach} & \textbf{Chan.} & \textbf{Tr.} & \textbf{Sepa-} & \textbf{Spk.} & \textbf{SDR} \\
 & \textbf{num.} & \textbf{data} & \textbf{ration} & \textbf{id.} & \textbf{[dB]} \\
 &  & \textbf{[hrs.]} & &  & \\ \hline
Mixture & - & - & - & - & 0.2 \\ \hline \hline
MCWF & 2 & - & Orac. & Orac. & 9.0 \\ \hline
MCWF & 4 & - & Orac. & Orac. & 13.4 \\ \hline\hline
TasNet~\cite{tasnet} & 2 & 50 & ML & ML & 8.4 \\ \hline
FD-SpkBeam~\cite{speakerbeam:timedomain} & 2 & 50 & ML & ML & 7.9 \\ \hline
TD-SpkBeam-Orig.~\cite{speakerbeam:timedomain} & 2 & 50 & ML & ML & 11.5 \\ \hline
TD-SpkBeam-Ext.\cite{mlse:spkbeamext1:icassp2021} & 2 & 50 & ML & ML & 12.9 \\ \hline
\hline
ILRMA~\cite{bss:ilrma:advsig2020} & 2 & - & BSS & Orac. & 5.9 \\ \hline
GCC-NMF~\cite{bss:nmf} & 2 & - & BSS & Orac. & 2.7 \\ \hline
MESSL~\cite{mandel2010} & 2 & - & BSS & Orac. & 3.3 \\ \hline
Prop. CSV$+\PilotO$ & 2 & - & BSE & Orac. & 5.4 \\ \hline
ILRMA + ML spk. ident. & 2 & - & BSS & ML & 5.4 \\ \hline
FS-IVE$+\PilotX+$defl. & 2 & - & BSE & ML & 4.5 \\ \hline
Prop. CSV$+\PilotX+$defl. & 2 & - & BSE & ML & 4.1 \\ \hline\hline
ILRMA~\cite{bss:ilrma:advsig2020} & 4 & - & BSS & Orac. & 7.6 \\ \hline
GLOSS~\cite{bss:smplx:taslp2020} & 4 & - & BSS & Orac. & 9.3 \\ \hline
Prop. CSV$+\PilotO$ & 4 & - & BSE & Orac. & 9.6 \\ \hline
ILRMA +ML spk. ident. & 4 & - & BSS & ML & 7.2 \\ \hline
FS-IVE$+\PilotX+$defl. & 4 & - & BSE & ML & 7.7 \\ \hline
Prop. CSV$+\PilotX+$defl. & 4 & - & BSE & ML & 7.8 \\ \hline\hline
Prop. CSV$+\PilotX$ & 4 & - & BSE & ML & 6.0 \\ \hline

\end{tabular}
\end{table}

These methods are evaluated in terms of SDR implemented in the BSS\_EVAL toolbox~\cite{bsseval}.  CSV-AuxIVE operates in the STFT domain with an FFT length of $1,000$, hop-size of $100$ (the shift of the FSMN network), and an applied Hamming window; the sampling frequency is $8$~kHz. The length of the CSV-AuxIVE block is $2$~seconds ($L_T=160$ frames). The demixing filters are initialized with a vector of ones, because the locations of the sources and the topology of the microphone array are unknown. The enrollment set contains $18$~speakers; the X-vectors are computed using unused sentences from the original WSJ0 dataset. The publicly available implementation\footnote{https://github.com/d-kitamura/ILRMA} of consistent ILRMA~\cite{bss:ilrma:advsig2020} is used. ILRMA (using $100$~iterations) and MCWF\footnote{Different results of MCWF reported in~\cite{mc-wsj0-2mix} are caused by a short STFT length of $256$ used there.} were applied in the STFT domain with window length of $1024$ and hop-size $512$ samples. The results for TasNet were taken over from~\cite{speakerbeam:timedomain}; for MESSL and GCC-NMF, they were found in~\cite{mc-wsj0-2mix}. The other results originate from their respective references.

Restricting the methods to two channels, the results presented in Table~\ref{tab:exp:wsj0} show that the ML-based spatial+spectral filtering outperforms the blind spatial filtering by a large margin. The supervised methods are even comparable to oracle MCWF using two/four microphones. This is possible due to the existence of a strictly matching training part of the MC-WSJ0-2mix dataset. In this setting, CSV-AuxIVE with $\PilotX$ and deflation outperforms MESSL and GCC-NMF, but is outperformed by ILRMA. 

The two-channel setting is, however, arguably unfair for blind methods relying solely on spatial diversity of the sources. Utilization of four channels increases the SDR for all blind methods. Using $\PilotX$, deflation and $4$~microphones, CSV-AuxIVE\footnote{We placed examples of the extraction on our web-page: \url{https://asap.ite.tul.cz/demos/blind-extraction-of-target-speech-source-guided-by-piloting-and-deflation/}} is comparable to ML-based FD-SpeakerBeam and outperforms blind ILRMA performing full separation followed by ML-based speaker identification. Using $\PilotO$, CSV-AuxIVE achieves results comparable to GLOSS. The results confirm that CSV-AuxIVE coincides with FS-IVE if the mixed sources are static. The best performance overall is achieved by the variants of TD-SpeakerBeam, which approach the oracle MCWF using $4$ channels. 

Concerning the benefits of deflation, the failures of $\PilotX$ (discussed in Section~\ref{sec:pilot:limits}), caused by a weak SOI activity and limited classification accuracy, deteriorate significantly the average SDR. Considering $4$~microphones, CSV-AuxIVE using $\PilotX$ yields SDR lower by $3.6$~dB compared to CSV-AuxIVE using $\PilotO$. The deflation partly alleviates this issue and increases the average SDR by $1.8$~dB. 

\section{Conclusions}
\label{sec:concl}
This manuscript presents a novel method for target speech extraction from realistic mixtures. It consists of a combination of blind extraction using CSV-AuxIVE method and data-driven identification of the SOI. Due to decoupling of the extraction and the identification, the training required by the method is simpler compared to fully data-driven approaches. Moreover, the proposed method is applicable to a wide variety of realistic extraction scenarios without any adaptation. The guidance of the blind technique towards the SOI is ensured through two techniques: the piloting and the successive deflation of the multi-source mixture. Evaluation of the proposed approach leads to the following conclusions: 1) The presented frame-wise SOI identification applied to mixtures exhibits accuracy of $67\%$ in highly reverberated and noisy scenarios ($T_{60}=600$~ms and $\text{SIR}=0$~dB). 2) This accuracy is sufficient to form an efficient pilot able to guide the extraction in most scenarios. However, the embedding-based piloting fails when the mixture contains a small number of frames where the SOI is dominant, such as when the activity of the SOI is short and has a low energy level. These cases can be remedied using successive deflation of the mixture along with the re-estimation of the SOI. 3) The proposed non-intrusive assessment of extraction quality can successfully be used as a decision mechanism to determine whether the deflation should be applied. It is strongly correlated with the objective/perceptual criteria used to evaluate quality of speech; the Pearson coefficient between PLDA score and STOI improvements reaches a value of $0.83$. 4) The procedure as a whole is language independent. The accuracy of the speaker identification deteriorates slightly for an unseen language, but this has negligible effect on the extraction. 5) The CSV-AuxIVE achieves more precise extraction compared to a blind block-wise static approach for mixtures of moving sources. On mixtures of static sources, the piloted CSV-AuxIVE is comparably accurate to competing blind approaches performing full separation followed by the ML-based/oracle speaker identification. In contrast to full separation approaches, CSV-AuxIVE does not require to know the number of speakers. 6) The proposed approach achieves a lower performance compared to the state-of-the-art machine learning-based algorithms, as observed on widely known CHiME-4 and MC-WSJ0-2mix datasets. On the other hand, it does not require any scenario-specific training data.

\bibliographystyle{IEEEtran}
\bibliography{article}

\vspace{-33mm}
\begin{IEEEbiography}[{\includegraphics[width=1in,height=1.25in,clip,keepaspectratio]{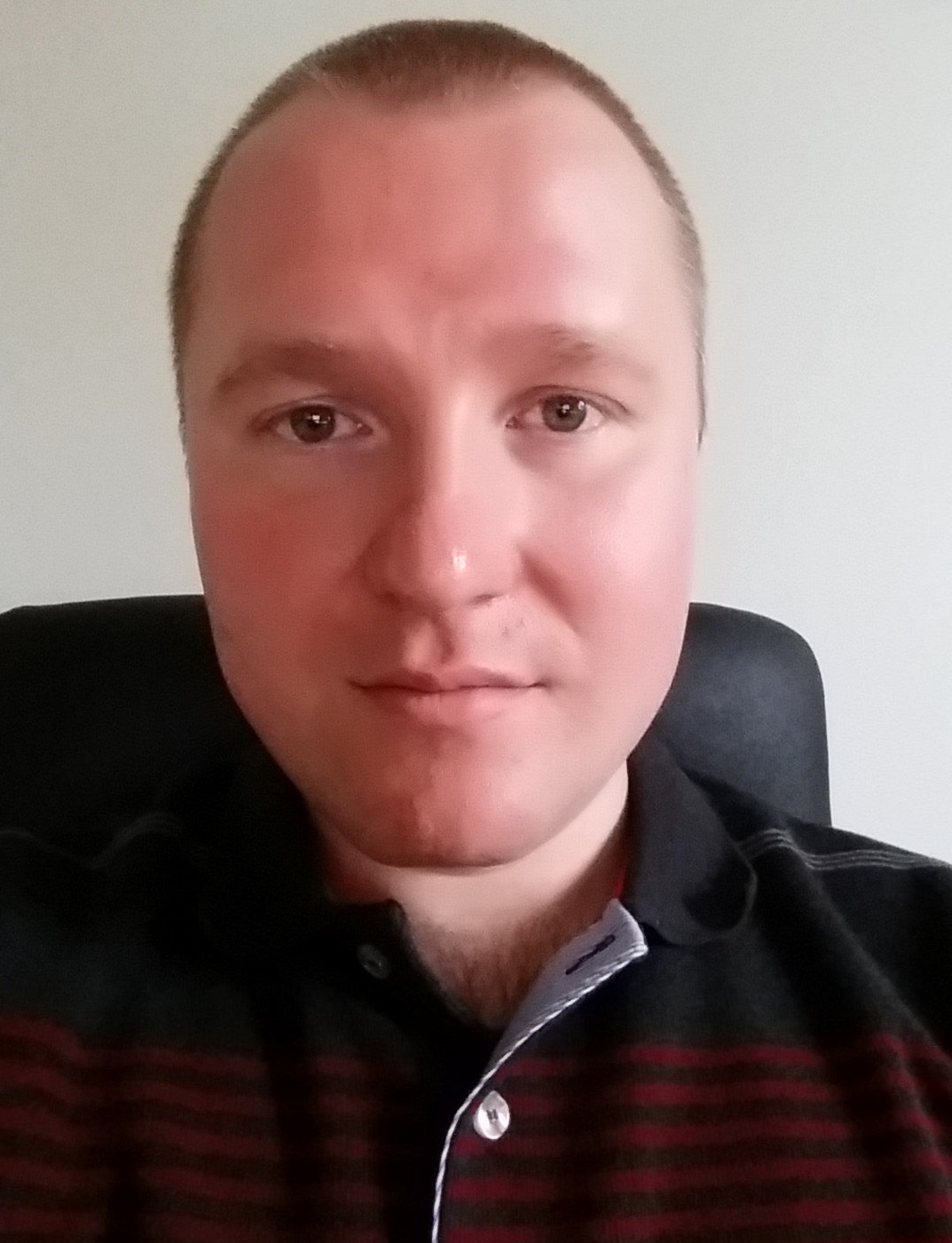}}]{Jiri Malek} (Member, IEEE) was born in Czechia in 1983. He received his Ph.D. in technical cybernetics from the Technical University of Liberec, Czechia, in 2011. Since 2011, he has been an Assistant Professor with the Faculty of Mechatronics, Technical University of Liberec. His main research interests include enhancement/separation of audio signals and robust automatic speech recognition. He is a reviewer for journals and conferences focused on digital signal processing, including the IEEE Transactions on Audio, Speech And Language Processing, IET Signal Processing, or ICASSP.
\end{IEEEbiography}

\vspace{-33mm}
\begin{IEEEbiography}[{\includegraphics[width=1in,height=1.25in,clip,keepaspectratio]{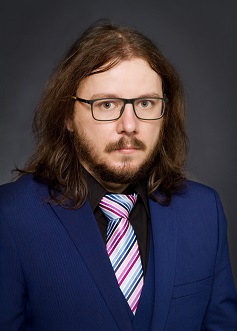}}]{Jakub Jansky} was born in Czechia, in 1989. He received his M.S. degree in application of software engineering from the Faculty of Nuclear Sciences and Physical Engineering, Czech Technical University in Prague, Czechia, in 2014. Since 2014, he has been a research assistant and Ph.D. student with the Faculty of Mechatronics, Technical University of Liberec. His main research interests include blind source separation, independent vector analysis, and sparse reconstruction.
\end{IEEEbiography}

\begin{IEEEbiography}[{\includegraphics[width=1in,height=1.25in,clip,keepaspectratio]{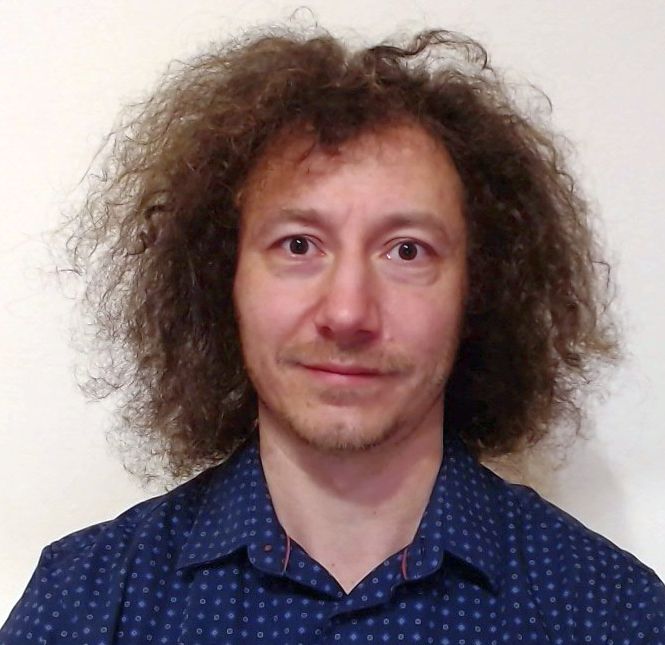}}]{Zbynek Koldovsky} (Senior member, IEEE) received the M.S. and Ph.D. degrees in mathematical modeling from the Faculty of Nuclear Sciences and Physical Engineering, Czech Technical University, Prague, Czech Republic, in 2002 and 2006, respectively. Since 2020, he has been a full professor with the Institute of Information Technology and Electronics,Technical University of Liberec, Liberec, Czech Republic, and the Leader of Acoustic Signal Analysis and Processing Group. He is currently the Associated Dean for Science, Research and Doctoral Studies with the Faculty of Mechatronics, Informatics and Interdisciplinary Studies. His main research interest is currently in blind source separation based on advanced mixing models applied in independent component/vector analysis and extraction. He has served as a General Co-Chair of the 12th Conference on Latent Variable Analysis and Signal Separation, Liberec, Czech Republic, and as a Technical Co-Chair of the 16th International Workshop on Acoustic Signal Enhancement, Tokyo, Japan. Since 2019, he has been a member of the IEEE SPS committee Audio and Acoustic Signal Processing. He served as the Area Chair for the area of Analysis of Speech and Audio Signals of Interspeech 2021 and 2022.
\end{IEEEbiography}

\vspace{-33mm}
\begin{IEEEbiography}[{\includegraphics[width=1in,height=1.25in,clip,keepaspectratio]{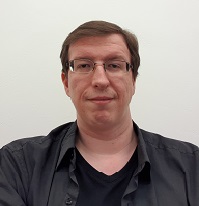}}]{Tomas Kounovsky} was born in Czechia in 1991. He received his M.S. degree in information technology from the Faculty of Mechatronics, Technical University of Liberec, in 2016. He is currently a Ph.D. student at the same faculty and university. His main research interests focus on audio signal processing, mainly speech enhancement and source separation.
\end{IEEEbiography}

\vspace{-33mm}
\begin{IEEEbiography}[{\includegraphics[width=1in,height=1.25in,clip,keepaspectratio]{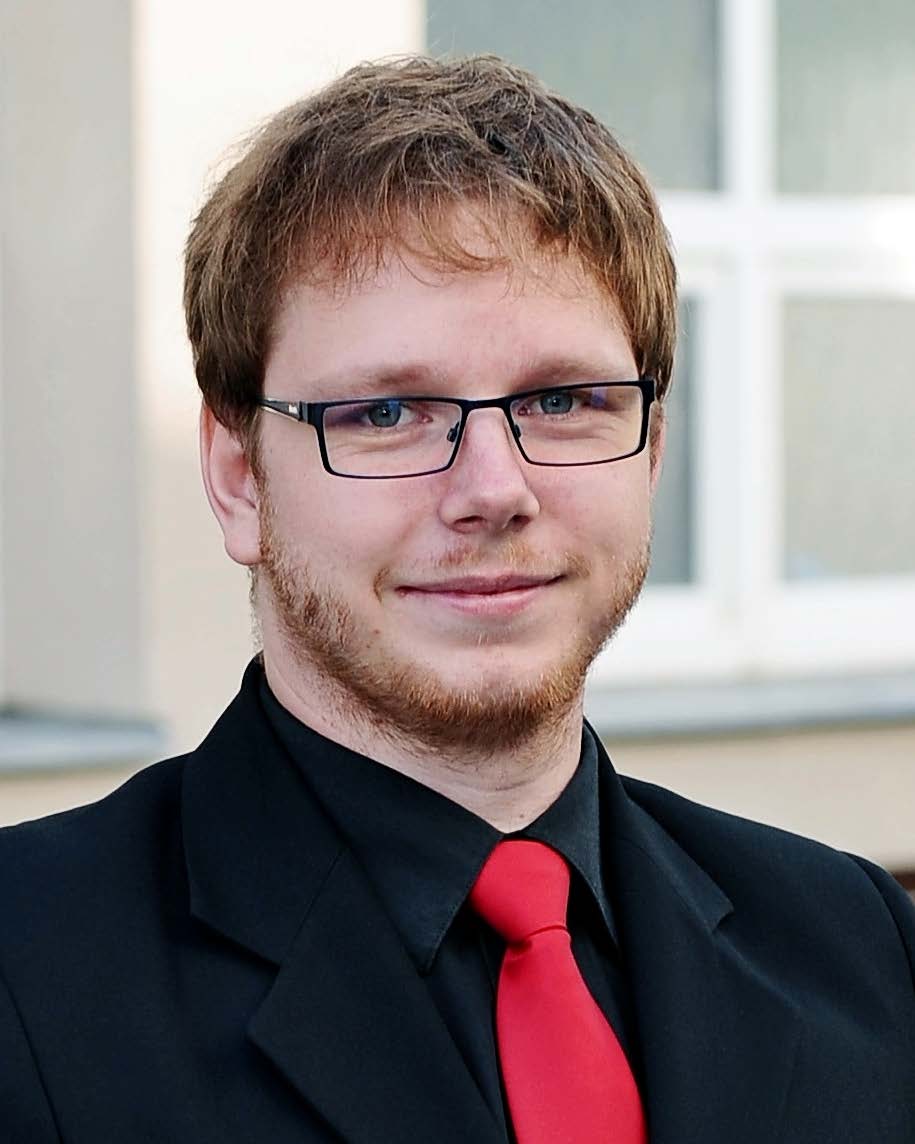}}]{Jaroslav Cmejla} received a master's degree in information technologies from the Faculty of Mechatronics, Informatics and Interdisciplinary Studies at the Technical University of Liberec in 2016. He is currently a Ph.D. student at the same faculty and university. He is a member of the Acoustic Signal Analysis and Processing Group led by prof. Zbynek Koldovsky. His main research interests are audio signal processing and blind source separation. His recent works are related to the blind source extraction problem.
\end{IEEEbiography}

\vspace{-33mm}
\begin{IEEEbiography}[{\includegraphics[width=1in,height=1.25in,clip,keepaspectratio]{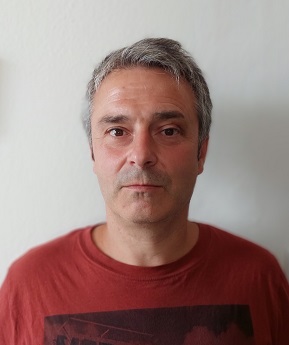}}]{Jindrich Zdansky} was born in Ceska Lipa, Czechia, in 1978. He received his M.S. degree in applied electronics from the Faculty of Electrical Engineering, Czech Technical University in Prague, Prague, Czechia, in 2002 and a Ph.D. in applied cybernetics from the Institute of Information Technology and Electronics, Technical University of Liberec, Liberec, Czechia 2006. Since 2005 member of the Speech Processing Group at the Technical University of Liberec. His main research interests include audio signal processing, voice-to-text, and speaker diarization technologies.
\end{IEEEbiography}

\end{document}